\documentclass[conference,compsoc]{IEEEtran}
\usepackage{tikz}

\usepackage{amssymb,amsmath,amsfonts,amsthm}
\newtheorem{theorem}{Theorem}
\usepackage{algpseudocode}
\usepackage{algorithm}
\usepackage{array}
\usepackage[caption=false,font=normalsize,labelfont=sf,textfont=sf]{subfig}
\usepackage{textcomp}
\usepackage{stfloats}
\usepackage{url}
\usepackage{verbatim}
\usepackage{graphicx}
\usepackage{cite}
\usepackage{soul}
\usepackage{subfig}
\usepackage{booktabs}  
\usepackage{multirow}  

\usepackage{caption}

\begin{document}
%
\title{NADD: Amplifying Noise for Effective Diffusion-based Adversarial Purification}

\author{\IEEEauthorblockN{David D. Nguyen}
\IEEEauthorblockA{CSIRO's Data61\\Australia}
\and
\IEEEauthorblockN{The-Anh Ta}
\IEEEauthorblockA{CSIRO's Data61\\Australia}
\and
\IEEEauthorblockN{Yansong Gao}
\IEEEauthorblockA{University of Western Australia\\Australia}
\and
\IEEEauthorblockN{Alsharif Abuadbba}
\IEEEauthorblockA{CSIRO's Data61\\Australia}
}


%


\maketitle

\pagestyle{plain}
\thispagestyle{plain}


\begin{abstract}
    The strategy of combining diffusion-based generative models with classifiers continues to demonstrate state-of-the-art performance on adversarial robustness benchmarks.
    Known as \textit{adversarial purification}, this exploits a diffusion model's capability of identifying high density regions in data distributions to purify adversarial perturbations from inputs.
    However, existing diffusion-based purification defenses are impractically slow and limited in robustness due to the low levels of noise used in the diffusion process.
    This low noise design aims to preserve the semantic features of the original input, thereby minimizing utility loss for benign inputs. 
    Our findings indicate that systematic amplification of noise throughout the diffusion process improves the robustness of adversarial purification.
    However, this approach presents a key challenge, as noise levels cannot be arbitrarily increased without risking distortion of the input.
    To address this key problem, we introduce high levels of noise during the forward process and propose the \textit{ring proximity correction} to gradually eliminate adversarial perturbations whilst closely preserving the original data sample.
    As a second contribution, we propose a new stochastic sampling method which introduces additional noise during the reverse diffusion process to dilute adversarial perturbations. 
    Without relying on gradient obfuscation, these contributions result in a new robustness accuracy record of 44.23\% on ImageNet using AutoAttack ($\ell_{\infty}=4/255$), an improvement of +2.07\% over the previous best work. 
    Furthermore, our method reduces inference time to 1.08 seconds per sample on ImageNet, a $47\times$ improvement over the existing state-of-the-art approach, making it far more practical for real-world defensive scenarios.
\end{abstract}

%
\IEEEpeerreviewmaketitle

\section{Introduction}

Deep neural networks continue to demonstrate remarkable progress across a broad range of problems. 
However, their weak robustness to adversarial perturbations \cite{szegedy2013intriguing} continues to hinder their potential in safety-first applications, such as medical imaging \cite{paschali2018generalizability,kaviani2022adversarial} or self-driving cars \cite{zhang2022adversarial,michaelis2019benchmarking}.
Although progress has been made in mitigating adversarial attacks, a complete solution remains elusive. 

One of the first proposals to improve the robustness of neural network classifiers was \textit{adversarial training} \cite{goodfellow2014explaining,madry2017towards}, which introduces adversarial perturbations into the training dataset. 
However, this approach often has difficulty generalizing effectively to previously unseen adversarial examples and attacks.
This requires retraining the entire neural network when new attacks are identified to maintain classifier robustness, which can be computationally expensive and impractical.

A promising method that addresses these limitations is \textit{denoised smoothing} \cite{salman2020denoised} which uses a \textit{denoiser} to directly remove adversarial perturbations from inputs. 
The protected classifier should accurately predict the label from the \textit{denoised input}, provided that the denoised samples are from the same distribution as the classifier's training data. 
More recently, the denoiser has been implemented using generative models, leading to a new line of defensive techniques referred to as \textit{adversarial purification} \cite{srinivasan2021robustifying,yoon2021adversarial,yang2019me}.

Denoising diffusion models \cite{sohl2015deep,ho2020denoising,song2020score,karras2022elucidating} are currently the state-of-the-art generative models for adversarial purification owing to their forward and reverse diffusion processes. 
The forward diffusion process dilutes adversarial perturbations by introducing noise into the data sample.
The reverse diffusion process simultaneously removes noise and push the samples towards higher density regions of the data distribution \cite{song2018pixeldefend,nie2022diffusion,carlini2022certified}.
This reduces the likelihood of misclassification, since higher-density regions are less likely to contain adversarial examples.

Our key finding is that systematic amplification and balancing of noise throughout the forward and reverse diffusion processes can significantly improve robustness against white-box attacks.
However, existing state-of-the-art purification frameworks \cite{nie2022diffusion,wang2022guided,lee2023robust,carlini2022certified} only utilize low levels of noise in the forward process in order to preserve the semantics of the original data sample. 
To introduce higher noise without decreasing classification accuracy, we propose the \textit{ring proximity condition}, which specifies an optimal region for denoised samples. 


Based on this analysis, we
introduce a new purification framework, \textbf{Noised-Amplified Diffusion Defense (NADD)}, that utilizes significantly higher amounts of noise in both the forward and reverse process. 
In particular, this framework introduces three novel techniques: (1) a \textit{ring proximity correction} step, (2) a correction schedule, and (3) a stochastic sampling method. 
We theoretically show that the purified sample will be bounded in a ring-shaped neighborhood of the original sample. 

In contrast to previous works that rely on Variance Preserving Stochastic Differential Equation (VPSDE), our NADD framework is built on top of EDM \cite{karras2022elucidating}, another diffusion model, resulting in much faster diffusion sampling time and reducing the number of time-steps from 1000 to less than 38.
Concretely, our method achieves an inference time of 1 second on ImageNet, which is $47\times$ faster than the previous best baseline \cite{lee2023robust}. This demonstrates the practicality of our method.

Overall, NADD achieves a state-of-the-art robustness accuracy of 44.23\% on ImageNet, which is a new record for diffusion-based adversarial purification \textit{and} adversarial training methods.
In summary, our key contributions are \footnote{Our source code and model weights will be released upon publication. We will also release a multi-node fork of AutoAttack (https://github.com/fra31/auto-attack) for the community.}:

\begin{itemize}
   
    \item We analyze the purification process and propose an optimal output region defined by the \textit{ring proximity condition}. This region is targeted during the reverse diffusion process to improve purification reconstruction quality and robustness. 

    \item  Based on the ring proximity condition, we propose a new purification framework called NADD. This framework can employ higher noise levels during the diffusion process to improve robustness against white-box attacks. 
        
    \item The NADD framework introduces three new techniques for purification defense: (1) ring proximity correction step, (2) a tailored correction schedule and (3) stochastic sampling. We theoretically justify that these techniques enable the purified samples to remain close to the original inputs, effectively removing adversarial noise without distorting data.
    
    \item We comprehensively evaluate our framework on a wide-range of gradient-based attacks and achieve  state-of-the-art robustness accuracy on CIFAR10 and ImageNet. 
\end{itemize}

These contributions improve the state of adversarial purification, making it both more robust and computationally feasible for practical deployment in safety-critical applications.

\section{Related Works}

Adversarial training \cite{goodfellow2014explaining,madry2017towards} is a fundamental technique that improves the robustness of neural network classifiers against adversarial attacks. 
This approach introduces adversarial examples into the training dataset, exposing the neural network to perturbations. 
Various extensions include adversarial data augmentation \cite{volpi2018generalizing,rebuffi2021data} and ensemble adversarial training \cite{tramer2017ensemble} which introduce additional data to improve robustness and generalization.
While adversarial training performs effectively against perturbations in the training set, it suffers against unseen attacks \cite{hendrycks2021unsolved,bai2021recent}.
Generating adversarial examples during training also significantly increases the overall training time, making it challenging for larger datasets, such as ImageNet \cite{xie2019intriguing}. 
However, the problem of generalization to new adversarial attacks without retraining remains largely unsolved, highlighting the need for complementary solutions such as adversarial purification.

Adversarial purification is an emerging family of adversarial defenses that aims to restore clean inputs from adversarially perturbed samples, typically using generative models, such as generative adversarial networks (GANs) \cite{samangouei2018defense}, auto-regressive models \cite{song2018pixeldefend}, energy-based models \cite{yoon2021adversarial} and, more recently, diffusion models \cite{nie2022diffusion}.
Unlike adversarial training, which modifies the classifier's learning process, adversarial purification employs the generative model as a pre-processing step that removes perturbations before classification on the purified input. 

Diffusion models have performed impressively due to their ability to handle high-dimensional data and progressively remove perturbations through their forward and reverse diffusion processes \cite{ho2020denoising,song2020score}.
Further improvements have been demonstrated by employing an ensemble of diffusion models \cite{xiao2023densepure}, incrementally introducing noise in multiple diffusion runs \cite{lee2023robust} and introducing a classifier to guide the diffusion model towards the correct class \cite{lin2024robust,zhang2024classifier}.

\begin{table}[t]
\centering
\caption{Comparison of diffusion-based purification methods. Inference time for ImageNet256 images using a single H100 GPU.}
\small  
\begin{tabular}{lcccc}
\toprule
Method & \makebox[0pt][c]{Diffusion}  & Discrete  & Number  & {Inference}\\
       &  \makebox[0pt][c]{Family}  &   Steps   & of Runs & {Time (s)} \\
\midrule
Nie et al. \cite{nie2022diffusion} & VPSDE & 100 & 1& 4.13 $\pm$ 0.14 \\
Lee et al. \cite{lee2023robust} & VPSDE & 620 & 8& 46.98 $\pm$ 0.25\\
Lee et al. \cite{lee2023robust} & VPSDE & 200 & 1& 15.88 $\pm$ 0.87 \\
Ours & EDM & \textbf{29} & 1 & \textbf{1.08 $\pm$ 0.02} \\
\bottomrule
\end{tabular}
\label{tbl:diffcompare}
\end{table}

Existing diffusion-based purification methods \cite{nie2022diffusion, lee2023robust} primarily rely on frameworks such as Variance Preserving Stochastic Differential Equations (VPSDE). 
While effective, these approaches face limitations due to their computational expense and large number of discrete steps as shown in Table~\ref{tbl:diffcompare}, rendering them impractical for deployment in large-scale applications. 
Furthermore, these methods often utilize low noise levels during the forward diffusion process to maintain reconstruction quality and preserve utility of benign inputs, which restrict their robustness when countering strong adversarial attacks. 
Additionally, the classifiers employed in current ``guidance" strategies are susceptible to adversarial example attacks, weakening the overall defense mechanism. 
To address these challenges, we propose a new correction technique that improves robustness in the presence of  higher forward diffusion noise levels and a more efficient sampling framework.



\section{Preliminaries} 
\noindent \textbf{Notation.}
We consider the classification setting, where there exists a data distribution $p_{data}$ with image samples represented as $\mathbf{x} \in \mathbb{R}^{C \times H \times W}$.
Each image sample is paired with a label $c$, typically a discrete index or a prompt. 
The goal of a classifier $f$ is to predict a label $\hat{c}$ given the data sample as input such that $f(\mathbf{x}) = \hat{{c}} \approx {c}$. 
\newline

\begin{figure*}[h]
    \centering
    \subfloat[\footnotesize Successful purification\label{fig1a:success}]{\includegraphics[width=0.48\textwidth]{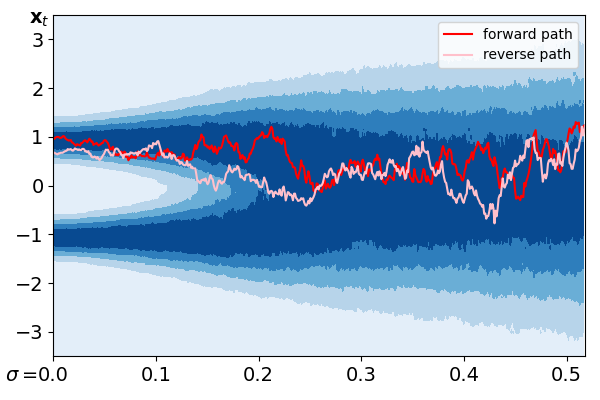}}\hspace{0.01\textwidth}
    \subfloat[\footnotesize Purification error\label{fig1b:error}]{\includegraphics[width=0.48\textwidth]{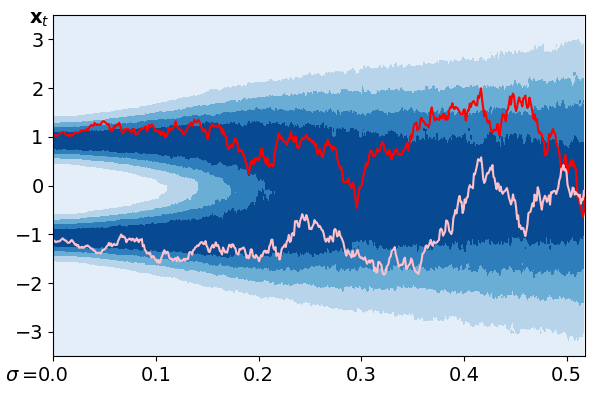}}

    \caption[]{ Diffusion trajectories for a bimodal data distribution according to a VPSDE. \subref{fig1a:success} The red line depicts the forward path, where noise is added to a sample evolving according to the forward diffusion process. The pink line shows the reverse path, ideally returning the noisy sample to its initial mode, centered at 1.  \subref{fig1b:error} An example of \textbf{purification error}, where a noisy sample from the mode centered at 1 follows a reverse path (orange) that incorrectly returns it to another mode, centered at -1. }%
    \label{fig:forward-backward-path}
\end{figure*}

\begin{figure*}[h]
    \centering
    \subfloat[\footnotesize Classifier Condition \label{fig2a:class}]{\includegraphics[width=0.32\textwidth]{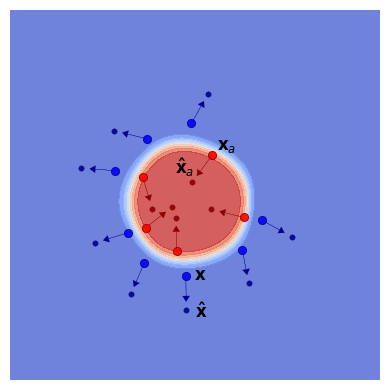}}\hspace{0.01\textwidth}
    \subfloat[\footnotesize  Proximity Condition\label{fig2b:prox}]{\includegraphics[width=0.32\textwidth]{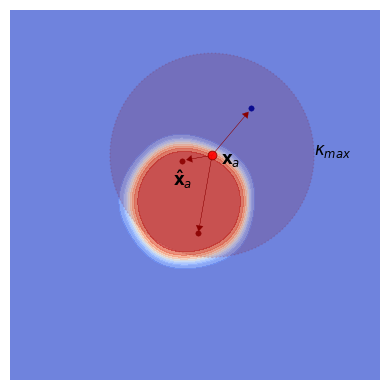}}
    \subfloat[\footnotesize  Ring Proximity Condition\label{fig2c:ours}]{\includegraphics[width=0.32\textwidth]{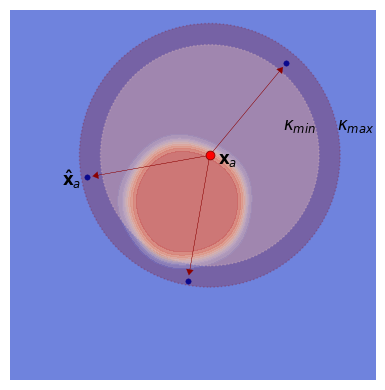}}

    \caption[]{Output space of the original image depicting the true class ({blue}) and adversarial class (red) regions. Adversarial examples $(\mathbf{x}_a)$ will sit near decision boundaries. \subref{fig2a:class} Class conditioned purification relies upon the guidance of a classifier. A compromised classifier will push purified adversarial examples  $(\hat{\mathbf{x}}_a)$ and benign examples $(\mathbf{\hat{x}})$ away from the decision boundary. \subref{fig2b:prox} Proximity conditioned purification bounds the purified samples within a local region (dark red) defined by $\kappa_{max}$. A significant portion of this local region will include the adversarial class. \subref{fig2c:ours} Ring proximity condition will concentrate purified samples within a ring-like region (dark red) defined by $\kappa_{min}$ and $\kappa_{max}$. This reduces the probability of producing a sample from the adversarial class.}
    \label{fig2:conditioncomparison}
\end{figure*}

\noindent \textbf{Adversarial Example Attacks.}
Adversarial examples \cite{szegedy2013intriguing, goodfellow2014explaining} are perturbations $\delta \in \mathbb{R}^{C \times H \times W}$ added to the data sample $\mathbf{x}_a = \mathbf{x} + \delta$ that cause the classifier to misclassify the label $f(\mathbf{x}_a) = \hat{{c}} \neq {c}$ without disrupting the semantics to human.
To measure the imperceptibility of each perturbation, vector norms such as the \( L_\infty \) or \( L_2 \) norm are commonly used.
These norms quantify the distance between the original sample \( \mathbf{x} \) and the adversarial sample \( \mathbf{x}_a \) in normalized [0,1] space. 
Specifically, the \( L_\infty \) norm bounds the maximum absolute change to any pixel by a value \( r \), while the \( L_2 \) norm constrains the perturbation to lie within a radius \( r > 0 \) from \( \mathbf{x} \) in Euclidean space.

A differentiable technique to generate adversarial examples is the Projected Gradient Descent (PGD)  \cite{madry2017towards} attack which assumes that the attacker can observe all information about the defender and classifier. 
PGD uses an iterative update rule to create an adversarial example $\mathbf{x}_a$:
\begin{equation} \mathbf{x}_{i+1} = \mathbf{x}_i + \alpha_i \  \text{sign}\left( \nabla_{\mathbf{x}_i} \mathcal{L}(f(\mathbf{x}_i), c) \right) 
\end{equation}
where $\alpha$ is the step size and $\mathcal{L}$ is cross-entropy loss.
For randomized defences, Expectation over Transformations (EOT) \cite{athalye2018synthesizing} is used to take an expectation over multiple random samples of the gradients which can improve PGD's success rate. 
For non-differentiable defences, attackers can use the Backward Pass Differentiable Approximation (BPDA) \cite{athalye2018obfuscated}, which provides an approximation of the non-differentiable function. 
We benchmark the robustness of our purification techniques against these attacks which aligns with previous works \cite{nie2022diffusion,lee2023robust}.
\newline 

\noindent \textbf{Denoised Smoothing.}
Denoised smoothing \cite{salman2020denoised} is a defensive technique that introduces a denoiser $D$ to directly remove an adverarial perturbation $\delta$:
\begin{equation}
    f(D(\mathbf{x} + \delta)) =  f(\hat{\mathbf{x}}) \approx f(\mathbf{x})
\end{equation}
The key assumption here is that the denoiser $D$ produces denoised samples $\hat{\textbf{x}}$ that are from the same data distribution as the original samples $\mathbf{x} \sim p_{data}$. 
Recent studies have proposed implementing the denoiser using a \textit{denoising diffusion model} \cite{carlini2022certified,nie2022diffusion}, resulting in a line of techniques known as \textit{adversarial purification}.
To best understand this, we first delve into diffusion models before discussing adversarial purification.
\newline

\noindent \textbf{Continuous-Time Diffusion Models.}
Denoising diffusion models \cite{sohl2015deep,song2020score,karras2022elucidating,ho2020denoising} are a family of generative models which consists of two processes: a forward process repeatedly adds noise to a data distribution $p_{data}(\mathbf{x})$, and a reverse process that generates new data samples from a tractable prior distribution $\pi(\mathbf{x})$, typically Gaussian. 

The forward direction of the diffusion process can be modelled by a stochastic differential equation (SDE) \cite{song2020score}: 
\begin{equation}
    \label{eq:forward-sde}
    \text{d}\mathbf{x}_t = \boldsymbol{\mu}(\mathbf{x}_t, t) \text{d}t + \sigma(t) \text{d}\textbf{w}_t
\end{equation}
where $t \in [0,T]$, $\boldsymbol{\mu}(\mathbf{x}_t, t)$ is the drift coefficient, $\sigma(t)$ is the diffusion coefficient and  $\{\textbf{w}_t\}_{t\in[0,T]}$ is Brownian motion.
This represents a forward trajectory of increasingly noisy samples $\{\mathbf{x}_t\}_{t\in[0,T]}$, where the starting samples $\mathbf{x}_0 \sim   p_{0}(\mathbf{x}) \equiv p_{data}(\mathbf{x})$ and the final samples $\mathbf{x}_T \sim p_T(\mathbf{x}) \equiv \pi(\mathbf{x})$.

The reverse direction starts with samples trivially drawn from the prior distribution and then removing noise according to a probability flow ordinary differential equation (PF ODE) \cite{song2020score,karras2022elucidating}:
\begin{equation}
    \label{eq:flow-ode}
    \text{d}\textbf{x}_t = [\boldsymbol{\mu}(\mathbf{x}_t,t) -  \sigma(t) {\nabla}_\mathbf{x} \log p_t(\mathbf{x}_t)  ] \text{d}t
\end{equation}
where ${\nabla}_\mathbf{x} \log p_t(\textbf{x}_t)$ 
represents a \textit{time-dependent score function}.
The score function represents the gradient of the log probability of the data distribution and is approximated with a denoiser function $D(\mathbf{x}_t; t) \approx \mathbf{x}_t + \sigma(t)^2 \cdot{\nabla}_\mathbf{x} \log p_t(\textbf{x}_t)$.
Following prior works \cite{karras2022elucidating,dockhorn2021score,song2023consistency}, the denoiser is implemented as:
\begin{equation}
    D_\theta(\mathbf{x}_t; t) = c_{skip}(t) \mathbf{x}_t + c_{out}(t)F_\theta(\mathbf{x}_t,t)
\end{equation}
where $c_{skip}$ is a skip connection, $c_{out}$ scales the output magnitude and $F_\theta$ is a neural network parameterized with weights $\theta$. 
Reverse sampling is stopped when the time $t$ is lower than a near-zero scalar $\varepsilon$.

The denoiser $D_\theta$
can be trained with a $L_2$ denoising error \cite{hyvarinen2005estimation,karras2022elucidating}:
\begin{equation}
    \label{eq:scorematching}
    \mathbb{E}_{t \sim p(t; 0,T)} \mathbb{E}_{\mathbf{x} \sim p_{data}}  \left\| D_{\theta}(\mathbf{x}_t; t) - \mathbf{x} \right\|^2_2
\end{equation}
where $p(t; 0,T)$ is a probability distribution over the interval $[0,T]$.
The denoising model provides a result that can be used to solve for a discretized approximation of Eq. \ref{eq:flow-ode} by stepping backwards in time along a sample's trajectory using an numerical ODE solver, such as the Euler \cite{song2020improved} or Heun \cite{karras2022elucidating} solvers. 
This results in a solution trajectory $\{\mathbf{\hat{x}}_t\}_{t\in[\varepsilon,T]}$, where $\hat{\mathbf{x}}_\varepsilon$ should be approximately from the data distribution. \\

\noindent {\textbf{Adversarial Purification.}
Diffusion-based adversarial purification targets the removal of adversarial perturbations by using both the forward and backward diffusion processes. 
Concurrently introduced by Carlini \textit{et al.} \cite{carlini2022certified} and Nie \textit{et al.} \cite{nie2022diffusion}, this framework can be summarized in two stages.
First, a data sample $\mathbf{x}_{0}$ is diffused along the forward process (Eq.~\ref{eq:forward-sde}) for time $t^* \in [0,T]$. 
This results in noisy sample $\mathbf{x}_{t^*}$ containing a pre-specified level of noise $\sigma_{t^*}$.
The rationale for adding noise is to dilute and eliminate any adversarial perturbations in the sample. 

Second, a ODE solver (Eq.~\ref{eq:flow-ode}) denoises the sample using a diffusion model $D_\theta$, resulting in a purified data sample $\mathbf{\hat{x}_\varepsilon}$ that can be viewed as being from $p_{data}(\mathbf{x})$ without adversarial noise.
This purified data sample is then passed to the classifier:
\begin{equation}
    f(D_\theta(\mathbf{x}_{t^*};t^*)) = f(\mathbf{\hat{x}_\varepsilon}) = \hat{c} \approx c
\end{equation}

A key hyper-parameter with diffusion-based purification is time \( t^* \), which controls the amount of noise in the forward process.
Theorem 3.1 by Nie \textit{et al.} \cite{nie2022diffusion} show that samples from the data distribution and adversarial distribution will converge as the amount of noise $ \sigma_{t}$ increases over the course of the forward process.
Given a well-trained diffusion model, this implies that the forward process eliminates adversarial perturbations and that purified samples from the reverse process will be from the true data distribution.

However, this theorem does not guarantee that the reconstructed sample will be from the same class as the original data sample. 
In fact, applying too much noise (i.e., a large \( \sigma_{t^*} \)) removes global semantics, increase reconstruction error and leads to higher misclassification rates of the purified data $\hat{\bf x}_{\varepsilon}$. 
We refer to this problem as utility trade-off or \textbf{purification error} $\hat{c} \neq c$, as seen in Figure~\ref{fig:forward-backward-path}.
This trade-off constrains the robustness of existing purification frameworks against stronger adversarial attacks.
In the following section, we propose a new purification framework that improves this critical trade-off between higher noise and reconstruction quality.

\begin{figure*}[h]
    \centering
    \includegraphics[width=0.95\textwidth]{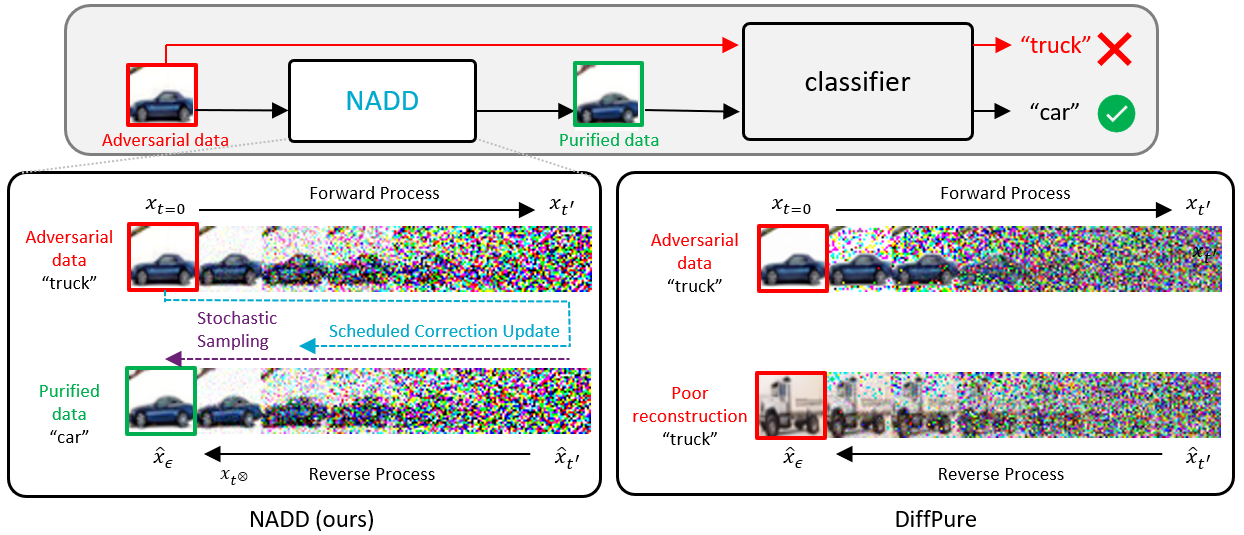}
    \caption[]{Comparison of purification frameworks for adversarial defense: \textbf{NADD} (left) and \textbf{DiffPure} (right). Both frameworks aim to restore adversarial inputs to their original, non-adversarial form before classification. In NADD, a high diffusion time during the forward process effectively removes adversarial perturbations. The \textbf{ring proximity corrections} ensures that purified data is close to the original and  \textbf{stochastic sampling} introduces noise throughout the reverse process. In contrast, DiffPure produces poor reconstructions semantically distant from the original input at high noise levels. }
    \label{fig:nadd-overview}
\end{figure*}

\section{Noise Amplified Diffusion Defence}
\label{sec:solution}


While previous diffusion-based purification approaches have to maintain a low amount of noise in the forward process to reduce purification error, we propose amplifying the noise to improve robustness against attacks.
Our conjecture is that increasing the amount of noise in defense plays an important role in deceiving white-box attacks that depend upon estimates from the diffusion model.

However, increasing noise blindly can destroy important semantic features within the data, adversely impacting the utility. 
We first analyze the purification procedure and propose the \textbf{ring proximity condition}, a property of the purified sample that should be satisfied to maintain low purification error in the presence of adversarial examples.
Following this, we introduce our new framework \textbf{Noise Amplified Diffusion Defence} (NADD) that satisfies this novel condition, by carefully introducing more noise in both the forward and reverse diffusion processes.
The pseudocode of NADD is provided in Algorithm~\ref{alg:backtracksampling}.

To enable faster sampling, we follow the theoretical diffusion framework of Karras \textit{et al.} \cite{karras2022elucidating} which discretizes time $T$ into $N-1$ time-steps, $\{t_i\}^N_{i=1}$ where $t_1=\epsilon$, $t_N=T$ and the noise schedule has a linear relationship to the time steps $\sigma(t)=t$. 
This diffusion framework significantly reduces the number of reverse time-steps which has the added benefit of full gradient back-propagation during adversarial robustness benchmarking. 

\begin{algorithm*}[t]
\caption{Noise Amplified Diffusion Defence (NADD)}
\label{alg:backtracksampling}
\begin{algorithmic}[1]
\Require $0 < t^\otimes	 < t^\prime \leq T$, $\sigma(t) = t$
\Function{Purify}{$\mathbf{x}$, $\mathbf{u}$, $\sigma$, $w$, $\gamma$, $t^\otimes	$, $t^\prime$}
\State $\mathbf{x}_{t_1} \gets \mathbf{x}$ \Comment{Start forward process with $\mathbf{x}_{t_0}$}
\State $i \gets 1$
\While{$t_i < t^\prime$} \Comment{Add excessive noise to data}
    \State $\mathbf{x}_{t_i} \gets \mathbf{x}_{t_{i-1}} + \mathcal{N}(\mathbf{0}, \mathbf{I} (\sigma({t_{i}})^2 - \sigma({t_{i-1}})^2))$ \Comment {Take forward step from $t_{i-1}$ to $t_{i}$}
    \State $i  \gets i + 1$
\EndWhile
\State  $\mathbf{\bar{x}}_{t_1} = \mathbf{x}_{t_1} + \mathbf{u}$ \Comment{Create target for ring correction update}
\State $\mathbf{\hat{x}}_{t^\prime} \gets \mathbf{{x}}_{t^\prime}$ \Comment{Start reverse process with $\mathbf{x}_{t^\prime}$}
\While{$\epsilon < \sigma(t_i)$} 
    \State $\mathbf{z} \sim \mathcal{N}(\mathbf{0}, S^2_{noise} \mathbf{I})$ \Comment{Stochastic sampling update}
    \State $\hat{t}_{i+1} = t_{i+1} ( 1 + \gamma_i )$
    \State ${\bf \hat{x}}^\prime_{t_{i+1}} :=  {\bf \hat{x}}_{t_{i+1}} + \sqrt{\hat{t}^2_{i+1} - t^2_{i+1}} \mathbf{z}$ 
    \State $d_i \gets \Phi(\mathbf{x}^\prime_{t_{i+1}} , t_{i+1} ; \theta)$ \Comment{Compute Euler or Heun update from $t_{i+1}$ to $t_i$}
    \If{${t^\otimes	} < {t_i}$} \Comment{Use ring correction step if $t > {t^\otimes	}$}
        \State $c_{i} \gets \frac{\mathbf{\bar{x}}_{t_1} - \mathbf{\hat{x}}_{t_{i+1}}}{t_{i} - t_{i+1}}$ \Comment{Compute slope from $\mathbf{\hat{x}}_{t_{i+1}}$ to $\mathbf{\bar{x}}_{t_1}$}
        \State $d_{i} \gets d_{i}(1-w_{i}) + c_{i}w_{i}$ \Comment{Apply weighting according to schedule}
    \EndIf
    \State $\mathbf{\hat{x}}_{t_i} \gets \mathbf{\hat{x}}_{t_{i+1}} + (t_i - t_{i+1}) d_i $ \Comment{Take reverse step from $t_{i+1}$ to $t_i$}
    \State $i \gets i - 1$
\EndWhile
\State \Return $\hat{\mathbf{x}}_{t_1}$ \Comment{Return reconstructed sample at $t_1$}

\EndFunction

\end{algorithmic}
\end{algorithm*}

\subsection{Requirements for Purification with More Noise}
In contrast to traditional generation, diffusion-based purification involves both the addition of noise and recovery of the original data during inference. 
In other words, the forward and reverse processes can be thought of as a \textit{single end-to-end process}.
Considering this perspective, we can leverage information from the forward process to optimize the reverse process in the presence of high noise. 
Below, we discuss several existing purification conditions, proposed by other studies, and then present our own, the \textit{ring proximity condition}.
\newline 

\noindent \textbf{Class Condition.} 
A desirable outcome of purification is that the reconstructed sample $\mathbf{\hat{x}}_{\varepsilon}$ belong to the same class as the original sample $\mathbf{x}_0$: 
 \begin{equation}
   f(\mathbf{x}_0) = f(\mathbf{\hat{x}}_{\varepsilon}) = c.
   \label{eq:class-condition}
\end{equation}
Ideally, this condition should be satisfied for benign samples, and is illustrated using a bimodal data distribution in Figure~\ref{fig1a:success}.
In this illustration, the sample follows a forward and reverse path that takes it back to its original mode. 
At increasingly higher noise levels, however, this condition is more difficult to satisfy. The reverse path of a noisy sample may drift into a different data region occupied by a different class, as illustrated in Figure~\ref{fig1b:error}.

An implementation that satisfies this condition is \textbf{classifier guidance} \cite{dhariwal2021diffusion,lin2024robust,zhang2024classifier}, where the diffusion model uses class predictions from a classifier to guide its solution trajectory towards a particular data region.
However, relying on classifier predictions prior to purification is highly risky where the original data samples are \textit{susceptible to adversarial perturbations}. 
As shown in Figure~\ref{fig2a:class}, we find that a misled classifier will push adversarial samples away from decision boundaries. 
\newline 

\noindent \textbf{Proximity Condition.} 
Without prior knowledge of the ground-truth class, we consider an auxiliary condition:
 \begin{equation}
    d(\mathbf{\hat{x}}_{\varepsilon}, \mathbf{x}_0) < \kappa
    \label{eq:condition2}
\end{equation}
where $d(\cdot,\cdot)$ is some distance metric and $\kappa$ defines the closeness between the purified sample $\mathbf{\hat{x}}_{\varepsilon}$ and original sample $\mathbf{x}_0$.
By controlling the proximity of $\hat{\mathbf{x}}_{\epsilon}$ to $\mathbf{x}_{0}$, according to $d$, we can indirectly satisfy Eq.~\ref{eq:class-condition} if we assume that the classifier behaves smoothly and doesn't change significantly within small local regions. 

A specific implementation of the proximity condition is \textit{GDMP} \cite{wang2022guided}, which uses the slope between $\mathbf{\hat{x}}_{t}$ and $\mathbf{x}_0$ to guide the diffusion model. 
However, this approach can cause the purified sample to reconstruct adversarial perturbations when the original sample is adversarial, $\mathbf{x}_0=\mathbf{x}_a$.
We observe that this issue arises because the purified sample is pushed \textit{too close} to the original adversarial sample.  
As a result, the method shows poor robustness under full gradient-based evaluation, as observed in Lee \textit{et al.} \cite{lee2023robust}.
\newline

\noindent \textbf{Ring Proximity Condition.} 
 To address the limitations of the proximity condition, we consider an improvement where the purified sample is sufficiently different from the original data sample to avoid reconstructing adversarial perturbations, yet remains close enough to maintain semantic class-based features:
\begin{equation}
\kappa_{\text{min}} < d(\mathbf{\hat{x}}_{\varepsilon}, \mathbf{x}_0) < \kappa_{\text{max}}.
\label{eq:modproxcondition}
\end{equation}

We refer to this objective as the \textit{ring proximity condition} which is implemented in our new purification framework, NADD.
This condition is implemented using a novel \textbf{ring proximity correction} (Sec. \ref{sec:correctionstep}) which effectively guides the diffusion model towards a region that sits between $k_{\min}$ and $k_{max}$.
Before we delve into these details, we first describe the theoretical framework that underlies NADD. 





\subsection{Theoretical Diffusion Framework}
The forward diffusion process plays a key role in introducing noise that eliminates adversarial perturbations. 
In existing purification frameworks, such as \textit{DiffPure} \cite{nie2022diffusion,lee2023robust}, the level of noise must be kept low to reduce reconstruction error between the original and purified sample, as seen in the r.h.s. of Figure~\ref{fig:nadd-overview}. 
In contrast, the NADD framework utilizes high amounts of noise to improve robustness against adversarial attacks while retaining the reconstruction error through new techniques: (1) ring proximity correction and (2) stochastic sampling. 
Below, we describe the underlying diffusion framework, followed by a description of these two techniques.\\

\noindent \textbf{Forward Process.} This framework introduces \textit{high level of noise} $\sigma_{t^\prime}$ to a data sample $\mathbf{x}_\varepsilon$ during the forward process, where $0 << t^\prime \leq T$. 
Each discrete step in the forward path produces $\mathbf{x}_{t_i}$ according to:
\begin{equation}\label{eq:forwar-step}
    \mathbf{x}_{t_i} := \mathbf{x}_{t_{i-1}} + \mathbf{z}, \ \text{where} \ \ \mathbf{z} \sim \mathcal{N}(\mathbf{0}, \mathbf{I} (\sigma({t_{i}})^2 - \sigma({t_{i-1}})^2)).
\end{equation}
The forward step is repeated until the Gaussian noise reaches $\sigma_{t^\prime}$, resulting in a forward trajectory $\{\mathbf{x}_t\}_{t\in[0,t^\prime]}$.
The final noisy sample $\mathbf{x}_{t^\prime}$ can be computed in closed form: $\mathbf{x}_{t^\prime} := \mathbf{x}_{t_{1}} + \mathbf{z}^\prime$ where $\mathbf{z}^\prime \sim \mathcal{N}(\mathbf{0}, \mathbf{I} (\sigma({t^\prime})^2))$. \\

\noindent \textbf{Reverse Process.} The final sample $\mathbf{x}_{t^\prime}=\mathbf{\hat{x}}_{t^\prime}$ is passed to a numerical ODE solver, which estimates the reverse trajectory $\{\mathbf{\hat{x}}_t\}_{t\in[\epsilon,t^\prime]}$.
A discretized reverse step of the solver updates $\mathbf{\hat{x}}_{t_{i+1}}$ to $\mathbf{\hat{x}}_{t_{i}}$ by:
\begin{equation}
    \mathbf{\hat{x}}_{t_i} := \mathbf{\hat{x}}_{t_{i+1}} + (t_i - t_{i+1}) \Phi(\mathbf{\hat{x}}_{t_{i+1}} , t_{i+1} ; \theta)
    \label{eq:fwddiscretestep}
\end{equation}
where $\Phi({\bf \hat{x}}_{t_{i+1}}, {t_{i+1}}; \theta)$ is an update function which can be implemented as the Euler or Heun solver. 
In the Euler case, the update function is $\Phi({\bf \hat{x}}_{t_{i+1}}, {t_{i+1}}; \theta)=(\mathbf{\hat{x}}_{t_{i+1}} -  D_\theta(\mathbf{\hat{x}}_{t_{i+1}};{t_{i+1}}))/t_{i+1}$.
For other solvers, such as Heun solver, $\Phi$ can be complicated, therefore we treat $\Phi$ as a black-box that employs a denoiser $D_\theta(\mathbf{x}_{t_i};t_i)$ to evaluate $\text{d}\mathbf{x}_{t_i} / \text{d}{t_i}$ at time-step $t_{i+1}$. 

\subsection{Ring Proximity Correction}
\label{sec:correctionstep}
\noindent \textbf{Correction Update.} To reduce reconstruction error caused by excessive forward noise, we propose a correction step that slopes towards a target sample $\mathbf{\bar{x}}_{0}$ at time $t_i$:
\begin{equation}
    c_{i} = \frac{\mathbf{\bar{x}}_{0} - \mathbf{\hat{x}}_{t_{i+1}}}{t_{i} - t_{i+1}}.
\end{equation}

The target sample is produced according to: 

 \begin{equation} \mathbf{\bar{x}}_{0} = \mathbf{x}_{0} + \mathbf{u}, 
 \end{equation} where: 
 \begin{equation} \mathbf{u} = r \frac{\mathbf{v}}{|\mathbf{v}|}, \quad r \sim \mathcal{U}[\kappa_{\text{min}}, \kappa_{\text{max}}], \quad \mathbf{v} \sim \mathcal{N}(0, I). \end{equation}

This ensures that the perturbation $\mathbf{u}$ lies on the surface of an $n$-dimensional sphere and is uniformly scaled by $r$, sampled from the range $[\kappa_{\text{min}}, \kappa_{\text{max}}]$. 
This construction satisfies the ring proximity condition described in Eq.~(\ref{eq:modproxcondition}).

The correction update $c_i$ is introduced into Eq.~(\ref{eq:fwddiscretestep}) as a mixture with the ODE solver's update:
\begin{equation}\label{eq:correction}
    \mathbf{\hat{x}}_{t_i} := \mathbf{\hat{x}}_{t_{i+1}} + (t_i - t_{i+1}) \big[\Phi(\mathbf{x}_{t_{i+1}} , t_{i+1} ; \theta)(1-w_i)  + c_{i}w_i \big]
\end{equation}
where the weight $w_i \in [0,1]$ modulates the strength of the correction. \\

\noindent \textbf{Correction Schedule.} 
Given that the score vector field, corresponding to the gradient of the log probability density, becomes decreasingly noisy at lower time, it follows that the correction weight should be reduced proportionally as time decreases.
As such, we propose the following power-law correction schedule:
\begin{equation}
w_i =
\begin{cases}
    \left( \frac{t_i - t_1}{t_N} \right)^\beta & \text{if } t_i > t^\otimes	 \\
0 & \text{otherwise}
\end{cases}
\end{equation}
where $\beta \in [0,1]$ controls the rate of decay of the weight over time, with higher values of $\beta$ leading to a steeper decay.

Importantly, the correction weight is set to zero for timesteps \(t\) less than \(t^\otimes	 \in [0, t^\prime]\), which omits the correction step as the reverse process approaches \(t = 0\). 
This omission can be thought of as an early stopping mechanism, thus we refer to $t^\otimes$ as \textit{time stop correction}.
This mechanism also helps prevent the reconstruction of adversarial perturbations, as we will empirically demonstrate in Section~\ref{sec:experiments}.


\subsection{Stochastic Sampling}
\label{sec:stocsampling}
Here, we describe stochastic sampling, where noise is injected into the data sample during each reverse step.
More precisely, at reverse step $i$, a stochastic update $\mathbf{d}_{t_i}$ is produced by passing a noisy sample from the previous step  $\hat{\bf x}^\prime_{t_{i+1}}$ into the update function $\Phi$:   
\begin{equation}
    \mathbf{d}_{t_i} := \Phi({\bf \hat{x}}^\prime_{t_{i+1}}, {t_{i+1}}; \theta)
\end{equation}

A noisy sample $\hat{\bf x}^\prime_{t_{i+1}}$ is generated as follows:
\begin{equation}
    {\bf \hat{x}}^\prime_{t_{i+1}} :=  {\bf \hat{x}}_{t_{i+1}} + \sqrt{\hat{t}^2_{i+1} - t^2_{i+1}} \mathbf{z}_{i+1} \ \text{where} \ \ \mathbf{z} \sim \mathcal{N}(\mathbf{0}, S^2_{noise} \mathbf{I})
\end{equation}
The amount of noise is governed by $\hat{t}_{i+1} = t_{i+1} ( 1 + \gamma_i )$ where the noise factor $\gamma$ is set according to:
\begin{equation}
\gamma_i =
\begin{cases}
\min \left(\frac{S_{\text{churn}}}{N}, \sqrt{2} - 1 \right) & \text{if } t_i \in [S_{\text{min}}, S_{\text{max}}] \\
0 & \text{otherwise}
\end{cases}
\end{equation}
The addition of noise follows leading stochastic samplers \cite{karras2022elucidating} which were introduced to correct errors introduced in the previous step. 
In our context, we argue that the additional noise during sampling improves robustness by removing adversarial perturbation, akin to the forward process, without significantly affecting sample quality.




\section{Theoretical Proofs}
We provide theoretical analysis for 
proving that with high probability the denoising with correction procedure defined in Eq. (\ref{eq:correction}) can indeed guide the backward diffusion process to samples in a neighborhood of the original input.

\begin{theorem}[Returning estimate for denoising with correction]
\label{thm:estimates}
    Given a pretrained diffusion model with denoiser $D_{\theta}$, number of steps $T$ and diffusion coefficient function $\sigma(t)$. Assume that $(i)$ $\sigma(t) = t$ and $t_{i+1} - t_i \leq \Delta \frac{T}{N}$, for some constant $\Delta$; $(ii)$ the diffusion model is well-trained so that the denoiser is given by $D_\theta({\bf x}_t; t) = \nabla_{ {\bf x} } \log p_t( {\bf x}_t) \cdot \sigma(t)^2 + {\bf x}_t$. 
    \begin{enumerate}
        \item Given an input data point ${\bf x}$, and a proximity upper-bound  $\kappa_{\text{max}}$, for any $\delta^* >0$, with the choice of the correction weight $w_i \geq 1 - \frac{\kappa_{\text{max}}}{2 \sqrt{ \log \frac{2N}{\delta^*}} \sqrt{2 \Delta} T}$,  we have 
    \begin{equation}
        {\sf Pr}[ \|{\bf x} - {\bf x}_{t^\otimes	} \|_2 \leq \kappa_{\text{max}} ] \geq 1 - \delta^*
    \end{equation}
        \item If the lower bound $\kappa_{\text{min}}$ is small enough: $\kappa_{min} < \frac{1}{2\sqrt{2\pi} N}$, then we can choose the weights $(w_i)_{i = 1..N}$ so that 

        \begin{equation}
        {\sf Pr}[ \|{\bf x} - {\bf x}_{t^\otimes	} \|_2 \geq \kappa_{\text{min}} ] \geq \frac{1}{2\sqrt{2\pi} N} 
    \end{equation}

    In particular, the denoiser is expected to return a sample ${\bf x}_{t^\otimes	}$ such that $ \|{\bf x} - {\bf x}_{t^\otimes	} \|_2 \geq \kappa_{\text{min}} $ in $5N$ runs.
    \end{enumerate}
     
\end{theorem}
The choice $\sigma(t) = t$ here is inline with our use of EDM models \cite{karras2022elucidating} in this paper. It is possible to obtain similar estimates for other choices of the schedule $\sigma$ with a change of variable to bring $\sigma$ to $\sigma(t) = t$ and adjust our calculations below.

\textit{Sketch of the proof.}
We write the forward diffusion process as ${\bf x} = {\bf x}_{t_0} \to $ ${\bf x}_{t_1} \to $ $\ldots$ $ \to {\bf x}_{t_N}$ and the reverse process as ${\bf x}_{t_0} \leftarrow $ ${\bf \hat{x}}_{t_1} \leftarrow$ $\ldots$ $ \leftarrow {\bf \hat{x}}_{t_N} = {\bf x}_{t_N}$.
Our proofs are based on two main ideas. Firstly, each forward step ${\bf x}_{t_i} \to $ ${\bf x}_{t_{i+1}}$ and reverse step ${\bf \hat{x}}_{t_i} \leftarrow $ ${\bf \hat{x}}_{t_{i+1}}$ is an update by adding a Gaussian vector. 
Thus, we can write ${\bf x}_{t_i} - {\bf \hat{x}}_{t_i}$ as a sum of ${\bf x}_{t_{i+1}} - {\bf \hat{x}}_{t_{i+1}}$ with Gaussian vectors, and use a backward induction argument from the index $i= N-1$ to $i=0$ to obtain upper and lower bound on $\|{\bf x}_{t_0} - {\bf \hat{x}}_{t_0}\|_2$.
Secondly, at each induction step we need to use special properties of Gaussian random variables: concentration and lower probabilities bound, addition and subtraction of Gaussian vectors are Gaussian vectors.  
We describe the main steps in the proof and refer to the Appendix.~\ref{appx:proofs} for further details.

     Using the definition of a forward update step and a denoising step, we obtain the following equation for ${\bf x}_{t_i} - {\bf \hat{x}}_{t_i}$: 
     \begin{align} \label{eq:thm1}
         {\bf x}_{t_i} - {\bf \hat{x}}_{t_i} 
        =
        \big[({\bf x}_{t_{i+1}} - {\bf \hat{x}}_{t_{i+1}}) - {\bf z}_i - {\bf z}_i' \big](1-w_i) 
    \end{align}
    where $ {\bf z}_i, {\bf z}_i' \sim \mathcal{N}(\mathbf{0}, \mathbf{I} (t_{i+1}^2 - t_{i}^2))$, and $w_i$ is the correction weight.

    \textit{Upper-bound by $\kappa_{ \text{max} }$}: To obtain upper probability bound by induction, we use a simple observation that if
    $
    {\sf Pr}[ X \leq \varepsilon ] \geq 1- \delta \text{ and }
    {\sf Pr}[ X' \leq \varepsilon' ] \geq 1 - \delta',
    $
    then 
    \begin{equation}\label{eq:thm3}
         {\sf Pr}[ X+X' \leq \varepsilon + \varepsilon'] \geq 1 - (\delta + \delta')
    \end{equation}
    
    From step $i+1$ to $i$, assume that we already have an estimate 
    \begin{equation}
        {\sf Pr} \Big[ \| {\bf x}_{t_{i+1}} - {\bf \hat{x}}_{t_{i+1}}\|_2 \leq \varepsilon_{i+1} \Big] \geq 1 - \delta_{i+1}
    \end{equation}
    Our goal in the induction step is to prove 
    \begin{equation}
        {\sf Pr}\Big[ \| {\bf x}_{t_{i}} - {\bf \hat{x}}_{t_{i}}\|_2 \leq \varepsilon_{i} \Big] \geq 1 - \delta_{i}
    \end{equation}
    for some $\varepsilon_i$, $\delta_i$ depends on $\varepsilon_{i+1}$, $\delta_{i+1}$, $t_1, t_{i+1}$ and $w_i$. 
    We use the Eq. (\ref{eq:thm1}) above. By concentration inequality for Gaussian random vectors, for any $\lambda_{i+1} \geq 0$, which can be chosen later, it holds that (see \cite[Section 7]{abramowitz-handbook}) 
    \begin{equation*}
        {\sf Pr}\Big[ \| {\bf z}_i + {\bf z}_i' \|_2 \leq \lambda_{i+1}\Big] \geq 1 - 2e^{ - \frac{\lambda_{i+1}^2}{4(-t_i^2 + t_{i+1}^2)} }
    \end{equation*}

    If $ \| {\bf x}_{t_{i+1}} - {\bf \hat{x}}_{t_{i+1}}\|_2 + \| {\bf z}_i + {\bf z}_i' \|_2  \leq  \varepsilon_{i+1} + \lambda_{i+1}$, then the triangle inequality implies that
    \begin{align}
    \| {\bf x}_{t_i} - {\bf \hat{x}}_{t_i} \|_2 
    & \leq
    \big[\| {\bf x}_{t_{i+1}} - {\bf \hat{x}}_{t_{i+1}}\|_2 + \| {\bf z}_i + {\bf z}_i' \|_2 \big](1-w_i) \notag \\
    & \leq (\varepsilon_{i+1} + \lambda_{i+1}) (1-w_i) \notag
    \end{align}
    from which we obtain the estimate 
    \begin{align*}
        & {\sf Pr} \Big[ \| {\bf x}_{t_i} - {\bf \hat{x}}_{t_i} \|_2  \leq (\varepsilon_{i+1} + \lambda_{i+1}) (1-w_i) \Big] 
        \\
        &\quad \geq 1 - \delta_{i+1} - 2e^{ - \frac{\lambda_{i+1}^2}{4(-t_i^2 + t_{i+1}^2)} }
    \end{align*}
    Let us choose 
    \begin{equation*}
        \varepsilon_i := (\varepsilon_{i+1} + \lambda_{i+1}) (1-w_i) \text{ and } \delta_i := \delta_{i+1} + 2e^{ - \frac{\lambda_{i+1}^2}{4(-t_i^2 + t_{i+1}^2)}}
    \end{equation*}

    It remains to make the choices for values of $\lambda_i$'s and $w = \text{min}_{i} \  w_i $ so that $\varepsilon_0 \leq \kappa_{max}$ and $\delta_0 \leq \delta^*$ in order to obtain the desired estimate 
    \begin{align}
       {\sf Pr} \Big[ \| {\bf x}_{t_0} - {\bf \hat{x}}_{t_0} \|_2 < \kappa_{max}
       \Big] & \geq 
       {\sf Pr}\Big[ \| {\bf x}_{t_0} - {\bf \hat{x}}_{t_0} \|_2 < \varepsilon_0
       \Big] \notag \\
       & \geq 1 - \delta_0 \geq 1 - \delta^* \notag
    \end{align}
    
    To this end, we will estimate the value of $\varepsilon_0$ and $\delta_0$ using their induction formulas. At the index $i = N$, we have $\varepsilon_N = 0$ and $\delta_N = 0$ as $ {\bf x}_{t_{N}} := {\bf \hat{x}}_{t_{N}} $.
   The formula for $\delta_0$ is
    \begin{equation}
        \delta_0 = 2 \sum_{i=0}^{N-1} e^{ - \frac{\lambda_{i+1}^2}{4 (-t_i^2 + t_{i+1}^2) } } 
    \end{equation}
    Choosing $\lambda_{i} : = 2 \lambda \sqrt{- t_{i}^2 + t_{i+1}^2}$ simplifies $\delta_0$ as
    \begin{equation}
        \delta_0 = 2 N e^{- \lambda^2}
    \end{equation}
    We see that $\delta_0 \leq \delta^*$ for any value of $\lambda$ such that $\lambda \geq \sqrt{ \log \frac{2N}{\delta^*}}$. Let us fix such a value of $\lambda$ and proceed to estimate $\varepsilon_0$.

    By using the induction formula for $\varepsilon_i$'s, we obtain the following expression for $\varepsilon_0$:
    \begin{equation}
        \varepsilon_0 = \lambda_N (1-w)^N + \lambda_{N-1} (1 - w)^{N-1} + \ldots + \lambda_{1} (1 - w)
    \end{equation}
    from which we can make crude estimates to obtain 
    \begin{equation*}
        \lambda_i \leq 2 \lambda \sqrt{2 \Delta} \frac{T}{N} \text{ and } (1 - w)^i \leq (1 - w)
    \end{equation*}
    which implies 
    \begin{equation}
        \varepsilon_0 \leq N \cdot 2 \lambda \sqrt{2 \Delta} \frac{T}{N} (1-w) = 2 \lambda \sqrt{2 \Delta} T (1 - w)
    \end{equation}
    In conclusion, we have $2 \lambda \sqrt{2 \Delta} T (1 - w) \leq \kappa_{max}$ for any choice of $w$ such that 
    \begin{equation}
        w \geq 1 - \frac{\kappa_{max}}{2 \lambda \sqrt{2 \Delta} T} = 1 - \frac{\kappa_{max}}{2 \sqrt{ \log \frac{2N}{\delta^*}} \sqrt{2 \Delta} T}
    \end{equation}
    with $\lambda = \sqrt{ \log \frac{2N}{\delta^*}}$.

    \textit{Lower-bound by $\kappa_{ \text{min} }$}:  Starting with the first denoising step, we have 
    \begin{align} 
         {\bf x}_{t_{N-1}} - {\bf \hat{x}}_{t_{N-1}} 
        =
        (- {\bf z}_{N-1} - {\bf z}_{N-1}' )(1-w_{N-1}) 
    \end{align}
    where $ {\bf z}_{N-1}, {\bf z}_{N-1}' \sim \mathcal{N}(\mathbf{0}, \mathbf{I} (t_{N}^2 - t_{N-1}^2))$. 
    It follows that ${\bf x}_{t_{N-1}} - {\bf \hat{x}}_{t_{N-1}} $ is a Gaussian vector sampled from $\mathcal{N}(\mathbf{0}, 2(1-w_{N-1}) \mathbf{I} (t_{N}^2 - t_{N-1}^2))$.

    We use the following standard lower bound estimate for Gaussian random variable $ X \sim \mathcal{N}(0,1)$ (see \cite[Section 7]{abramowitz-handbook})
    \begin{equation}
        {\sf Pr}[|X| > \lambda] > \frac{x}{x^2 + 1} \frac{e^{-x^2/2}}{\sqrt{2\pi}}, \; \forall x>0
    \end{equation}
    By scaling and looking at only one coordinate of the Gaussian vector ${\bf x}_{t_{N-1}} - {\bf \hat{x}}_{t_{N-1}} $, we have a loose estimate
    \begin{equation*}
        {\sf Pr}[\|{\bf x}_{t_{N-1}} - {\bf \hat{x}}_{t_{N-1}}\|_2  > \lambda_{N-1}] > \delta_{N-1} := \frac{x}{x^2 + 1} \frac{e^{-x^2/2}}{\sqrt{2\pi}} 
    \end{equation*}
    for $ x = \frac{\lambda_{N-1}}{(2(1 - w_{N-1})(t_{N}^2 - t_{N-1}^2))^{1/2}}$.

    For induction, we use the following observation: if two random vectors ${\bf x}$ and ${\bf y}$ satisfy lower probability bounds ${\sf Pr}[|{\bf x} > \lambda|]$ and ${\sf Pr}[|{\bf y} > \lambda'+\lambda|]$, for some $\lambda, \lambda' > 0$,  
    then, it holds  
    \begin{align*}
        &{\sf Pr}[ \|{\bf x} - {\bf y}\|_2 > \lambda' ] 
        > {\sf Pr}[ \|{\bf x}\|_2 > \lambda] \cdot {\sf Pr}[ \lambda' + \lambda >  \|{\bf y}\|_2 ]
    \end{align*}

    We apply this estimate to our case which corresponds to ${\bf x} = ({\bf x}_{t_{i}} - {\bf \hat{x}}_{t_{i}})$, $\lambda = \lambda_i$, and ${\bf y} = ({\bf z}_{i} + {\bf z}_{i}')$, $\lambda' = \lambda_{i-1}(1 - w_{i-1})$, where $\lambda_i, \lambda_{i-1}$ will be chosen later. 
    The induction assumption is
    \begin{equation*}
        {\sf Pr}[\|{\bf x}_{t_{i}} - {\bf \hat{x}}_{t_{i}}\|_2 > \lambda_i]
        > \delta_i
    \end{equation*}
    from which we have 
    \begin{align*}
        &{\sf Pr}[ \|{\bf x}_{t_{i-1}} - {\bf \hat{x}}_{t_{i-1}}\|_2 > \lambda_{i-1} ] 
        \\
        &\quad > \delta_i \cdot (1 - 2e^{ - \frac{(\lambda_{i-1}(1 - w_{i-1}) + \lambda_i)^2}{4(-t_i^2 + t_{i+1}^2)} })
    \end{align*}
    We can choose 
    \begin{equation}
        \delta_{i-1} := \delta_i \cdot (1 - 2e^{ - \frac{(\lambda_{i-1}(1 - w_{i-1}) + \lambda_i)^2}{4(-t_i^2 + t_{i+1}^2)} })
    \end{equation}
    to obtain the following estimate for the induction step
    \begin{equation}
         {\sf Pr}[ \|{\bf x}_{t_{i-1}} - {\bf \hat{x}}_{t_{i-1}}\|_2 > \lambda_{i-1} ] > \delta_{i-1}.
    \end{equation}

    Recall that our goal is at step $i=0$, we have ${\sf Pr} [ \| {\bf x}_{t_0} - {\bf \hat{x}}_{t_0} \|_2 > \kappa_{min} ] > \delta_*$. This means $\lambda_0 = \kappa_{min}$ and $\delta_* = \delta_0$.

    Given $\kappa_{min}$ and $(t_i)_{i = 1..N}$, we choose $\lambda_0 = \kappa_{min}$ and $w_i-1, \lambda_i, i = 1, \ldots, N-1$ so that  
    \[
    1 - 2e^{ - \frac{(\lambda_{i-1}(1 - w_{i-1}) + \lambda_i)^2}{4(-t_i^2 + t_{i+1}^2)} } = 1- \frac{1}{i+1} = \frac{i}{i+1} 
    \]

    Then, we have 
    \begin{align*}
        \delta_0 & = \delta_{N-1} \prod_{i=1}^{N-1} (1 - 2e^{ - \frac{(\lambda_{i-1}(1 - w_{i-1}) + \lambda_i)^2}{4(-t_i^2 + t_{i+1}^2)} })
        \\
        & = \delta_{N-1} \prod_{i=1}^{N-1} \frac{i}{i+1} = \frac{\delta_{N-1}}{N} 
    \end{align*}

    Finally, we still have a free parameter $w_{N-1} < 1$ to choose so that $\delta_{N-1} \to \frac{1}{2\sqrt{2\pi}}$, for which $\kappa_{min} \to \frac{1}{2\sqrt{2\pi} N}$.





\section{Experiments}
\label{sec:experiments}

In this section, we first outline our experimental setup in Sec.~\ref{sec:expsetup} and then benchmark our method on various strong adversarial attack benchmarks against state-of-the-art adversarial training and adversarial purification frameworks (Sec.~\ref{sec:comparison}). 
We present ablation studies in Sec.~\ref{sec:ablation}, which provide more insights into our new framework. 




\begin{figure*}[h]
    \centering
    \includegraphics[width=\textwidth]{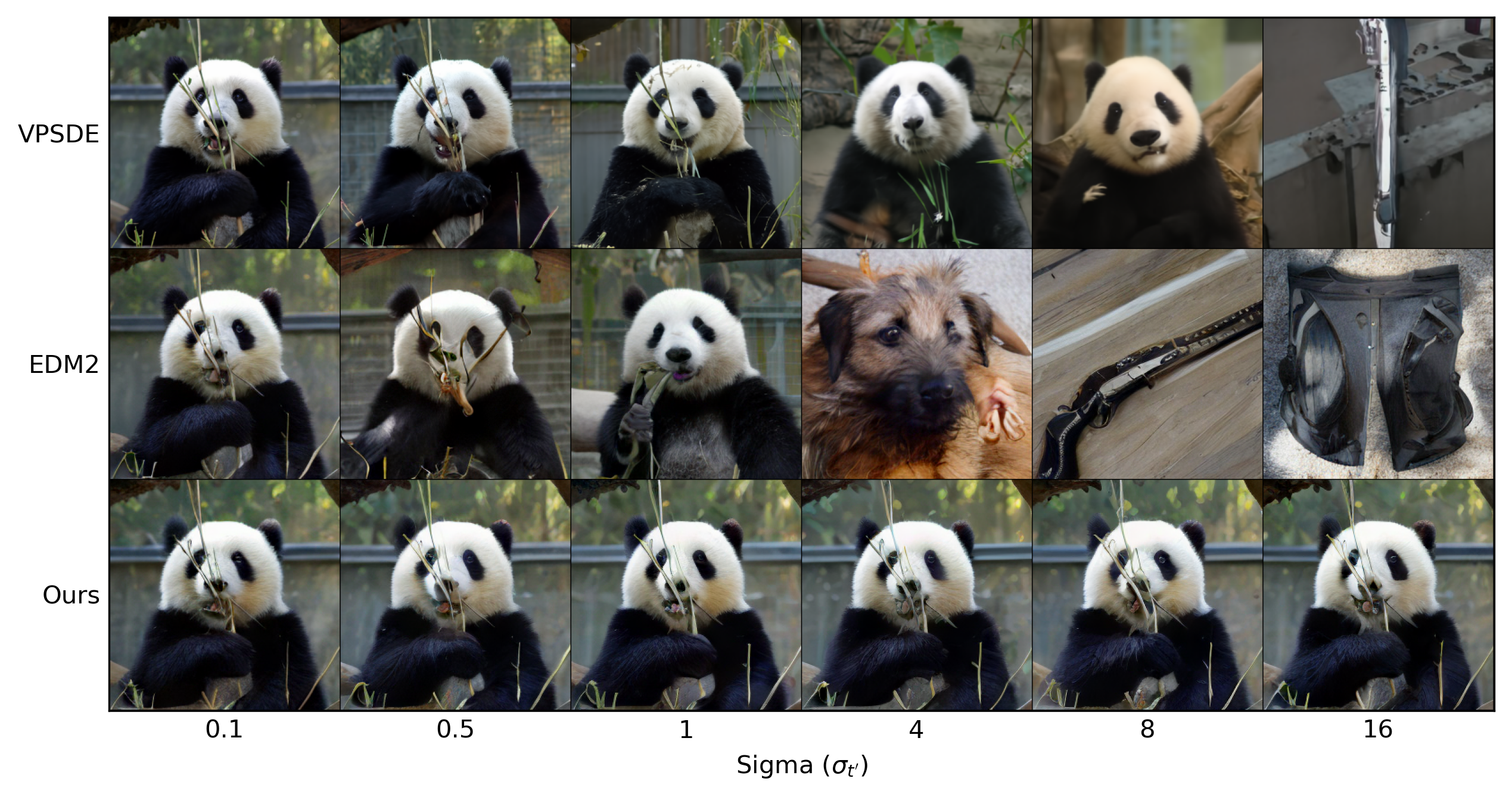}

    \caption[]{ImageNet reconstructions by three models at various noise levels defined by $\sigma_{t^\prime}$. The VPSDE model, as used by Nie et al. \cite{nie2022diffusion} and Lee et al. \cite{lee2023robust}, shows difficulty in preserving fine details, such as the bamboo stick, starting from $\sigma_{t^\prime} \geq 4$ and fails to maintain the recognizable structure of the panda at $\sigma_{t^\prime} = 16$. The EDM2 model \cite{karras2022elucidating} encounters similar challenges even at lower noise levels, displaying degradation in image quality. In contrast, our proposed NADD framework significantly enhances the EDM2 model's ability to reconstruct images, demonstrating consistent high-quality outputs even at high noise levels up to $\sigma_{t^\prime} = 16$.}
    \label{fig:imgnethighnoise}
\end{figure*}

\subsection{Setup}
\label{sec:expsetup}

\noindent \textbf{Datasets and Classifiers.}
For evaluation, we utilize two datasets: CIFAR10 (32x32) \cite{krizhevsky2009learning} and ImageNet (256x256) \cite{deng2009imagenet}, along with three pre-trained base classifiers: ResNet-50 \cite{he2016deep}, WideResNet-28-10 and WideResNet-70-16 \cite{zagoruyko2016wide}. 
This aligns with previous works \cite{nie2022diffusion,lee2023robust}. \\

\noindent \textbf{Baselines and Diffusion Models.}
Benchmarking diffusion models against optimization-based attacks is challenging as the reverse process can take up to several hundred time-steps to complete.
This requires significant computational resources to compute the full gradients of the diffusion model.

GDMP \cite{wang2022guided} masks the gradients of the diffusion model, however, this is not a realistic assumption under white-box attack.
The \textbf{adjoint method}, proposed by Nie \textit{et al.} \cite{nie2022diffusion}, provides an efficient way to compute gradients of the ODE solver, however, the approximations mean that the gradients are not exact. 
As pointed out by Lee \textit{et al.} \cite{lee2023robust} (in Table 1 of that work), the adjoint method can \textit{overstate} robust accuracy results. 
The most precise gradient approximation technique is the \textbf{surrogate process} \cite{lee2023robust}, which simply runs the reverse diffusion process with fewer and longer discrete time-steps. 
We use this approach where \textbf{full gradients} cannot be computed. 

We compare our technique against three state-of-the-art diffusion purification models by Yoon \textit{et al.} \cite{yoon2021adversarial}, Nie \textit{et al.}, \cite{nie2022diffusion}, and Lee \textit{et al.} \cite{lee2023robust}.
We report numbers directly from their respective papers.
We note that Lee \textit{et al.} technique requires 8 runs of the diffusion model while other baselines and our method only takes 1 run.
Furthermore, Lee \textit{et al.} employs different models during attack and defense, whereas others use the same model for both. 

In our approach, we utilize EDM \cite{karras2022elucidating} for CIFAR10 and EDM2 \cite{karras2024analyzing} for ImageNet with $T=38$ which allows for full gradient computations during benchmarking. Further hyperparameters can be found in the Appendix.\\

\noindent \textbf{Adversarial Example Attacks.}
Our evaluation framework follows the approach of Lee \textit{et al.} \cite{lee2023robust}, who proposes using both PGD+EOT and RobustBench \cite{croce2022evaluating} to benchmark diffusion-based purification. 
PGD+EOT uses 200 update iterations for CIFAR10 and 20 update iterations for Imagenet. 
We compare against adversarial training techniques using the $\ell_\infty$ and $\ell_2$ norm settings. 
As our diffusion-based defence employs stochasticity in the forward and reverse process, the adaptive attacks use Expectation Over Time (EOT) \cite{athalye2018obfuscated} with EOT=20. 
We also compare with existing adversarial purification methods using the BPDA+EOT attack \cite{hillstochastic}. 
\newline

\noindent \textbf{Metrics.}
We assess the effectiveness of defense methods using two key metrics: standard accuracy and robust accuracy. 
Standard accuracy reflects the model's performance on clean, unperturbed data and is evaluated across the entire test set of each dataset. 
Robust accuracy, on the other hand, indicates performance against adversarial examples crafted using adaptive attacks.
Given the high computational demands associated with adaptive attacks, we report robust accuracy on a fixed subset of 512 images, randomly selected from the test set, for both our approaches.
As shown by previous authors, there is not a significant difference between the sampled subset and whole test set \cite{nie2022diffusion,yoon2021adversarial,lee2023robust}.

\begin{table*}[t]
\centering
\caption{Standard and robust accuracy against PGD and AutoAttack on CIFAR-10 using $\ell_{\infty}$ ($\varepsilon=8/255$) and $\ell_{2}$ ($\varepsilon=0.5$) norm settings with various defences. The first three rows in each classifier group are adversarial training (AT), while the bottom three are adversarial purification (AP) methods. $^\ast$Extra data used. $^\ddagger$Eight diffusion cycles used.}
\small
\begin{tabular}{cc}
\begin{minipage}{.48\linewidth}
\centering
\begin{tabular}{lccc}
\toprule
\multirow{2}{*}{Method} & \multicolumn{3}{c}{Accuracy under $\ell_{\infty}$ Norm} \\
\cmidrule(lr){2-4}
& Standard & PGD & AutoAttack \\
\midrule
\multicolumn{4}{l}{\textbf{WRN-28-10}} \\
\midrule
Pang et al. \cite{pang2022robustness} & 88.62 & 64.95 & 61.04 \\
Gowal et al. \cite{gowal2020uncovering}$^\ast$ & 88.54 & 65.93 & 62.76 \\
Gowal et al. \cite{gowal2021improving} & 87.51 & 66.01 & 63.38 \\
\midrule
Yoon et al. \cite{yoon2021adversarial} & 85.66$\pm$0.51 & 33.48$\pm$0.86 & 59.53$\pm$0.87 \\
Nie et al. \cite{nie2022diffusion} & 90.07$\pm$0.97 & 46.84$\pm$1.44 & 63.06$\pm$0.81 \\
Lee et al. \cite{lee2023robust}$^\ddagger$ & 90.16$\pm$0.64 & 55.82$\pm$0.59 & 70.47$\pm$1.53 \\
Ours & \textbf{90.22$\pm$0.69} & \textbf{59.52$\pm$1.37} & \textbf{71.09$\pm$0.71} \\
\midrule
\multicolumn{4}{l}{\textbf{WRN-70-16}} \\
\midrule
Gowal et al. \cite{gowal2020uncovering}$^\ast$ & 91.10 & 68.66 & 65.87 \\
Gowal et al. \cite{gowal2021improving}& 88.75 & 69.03 & 66.10 \\
Rebuffi et al. \cite{rebuffi2021fixing}$^\ast$ & 92.22 & 69.97 & 66.56 \\
\midrule
Yoon et al. \cite{yoon2021adversarial} & 86.76$\pm$1.15 & 37.11$\pm$1.35 & 60.86$\pm$0.56 \\
Nie et al. \cite{nie2022diffusion} & 90.43$\pm$0.60 & 51.13$\pm$0.87 & 66.06$\pm$1.17 \\
Lee et al. \cite{lee2023robust}$^\ddagger$ & {90.53$\pm$0.14} & 56.88$\pm$1.06 & 70.31$\pm$0.62 \\
Ours & \textbf{90.68$\pm$0.84} & \textbf{60.01$\pm$1.91} & \textbf{70.98$\pm$0.75}\\
\bottomrule
\end{tabular}
\end{minipage} &
\begin{minipage}{.48\linewidth}
\centering
\begin{tabular}{lccc}
\toprule
\multirow{2}{*}{Method} & \multicolumn{3}{c}{Accuracy under $\ell_{2}$ Norm} \\
\cmidrule(lr){2-4}
& Standard & PGD & AutoAttack \\
\midrule
\multicolumn{4}{l}{\textbf{WRN-28-10}} \\
\midrule
Sehwag et al. \cite{sehwag2021robust} & 90.93 & 83.75 & 77.24 \\
Augustin et al. \cite{augustin2020adversarial} & 93.96 & 86.14 & 78.79 \\
Rebuffi et al.\cite{rebuffi2021fixing}$\ast$  & 91.79 & 85.05 & 78.80 \\
\midrule
Yoon et al. \cite{yoon2021adversarial} & 85.66$\pm$0.51 & 73.32$\pm$0.76 & 79.57$\pm$0.38 \\
Nie et al. \cite{nie2022diffusion} & 91.41$\pm$1.00 & 79.45$\pm$1.16 & 81.70$\pm$0.84 \\
Lee et al. \cite{lee2023robust}$^\ddagger$ & 90.16$\pm$0.64 & 83.59$\pm$0.88 & \textbf{86.48$\pm$0.38} \\
Ours & \textbf{93.26}$\pm$0.25 & \textbf{85.30$\pm$0.46} &  81.03$\pm$0.65 \\
\midrule
\multicolumn{4}{l}{\textbf{WRN-70-16}} \\
\midrule
Rebuffi et al. \cite{rebuffi2021fixing}  & 92.41 & 86.24 & 80.42 \\
Gowal et al.\cite{gowal2020uncovering}$\ast$ & 94.74 & 88.18 & 80.53 \\
Rebuffi et al.\cite{rebuffi2021fixing}$\ast$  & 95.74 & 89.62 & 82.32 \\
\midrule
Yoon et al. \cite{yoon2021adversarial} & 86.76$\pm$1.15 & 75.66$\pm$1.29 & 80.43$\pm$0.42 \\
Nie et al. \cite{nie2022diffusion} & 92.15$\pm$0.72 & 82.97$\pm$1.38 & 83.06$\pm$1.27 \\
Lee et al. \cite{lee2023robust}$^\ddagger$ & 90.53$\pm$0.14 & 83.75$\pm$0.99 & \textbf{85.59$\pm$0.61} \\
Ours & \textbf{93.21$\pm$0.56} & \textbf{85.15$\pm$0.69} & 82.15$\pm$0.85 \\
\bottomrule
\end{tabular}
\end{minipage}
\end{tabular}
\label{tbl:combined-cifar10}
\end{table*}

\begin{table}[ht]
\centering
\caption{Standard and robust accuracy against BPDA+EOT $\ell_{\infty} (\varepsilon = 8/255)$ on CIFAR-10 using WideResNet 28-10. $^\dagger$One diffusion cycle used.}
\small 
\begin{tabular}{lllc}
\toprule
\multirow{2}{*}{Method} & \multirow{2}{*}{Technique} & \multicolumn{2}{c}{Accuracy (\%)} \\
\cmidrule(lr){3-4}
& & Standard & Robust \\
\midrule
Song et al. \cite{song2018pixeldefend} & Gibbs Update & 95.00 & 9.00 \\
Yang et al. \cite{yang2019me} & Mask+Recon & 94.00 & 15.00 \\
Hill et al. \cite{hill2020stochastic} & EBM+LD & 84.12 & 54.90 \\
\midrule
Yoon et al. \cite{yoon2021adversarial}  & DSM+LD & 85.66$\pm$0.51 & 66.91$\pm$1.75 \\
Nie et al. \cite{nie2022diffusion} & DiffPure+VPSDE & \textbf{90.07$\pm$0.97} & 81.45$\pm$1.51 \\
Lee et al. \cite{lee2023robust}$^\dagger$ & DiffPure+VPSDE & 89.67$\pm$1.54 & {82.31$\pm$2.10} \\
Ours  & NADD+EDM & 89.76$\pm$0.87 & \textbf{85.24$\pm$0.95} \\
\bottomrule
\end{tabular}
\label{tbl:bpda}
\end{table}

\subsection{Comparison to State-of-The-Art}
\label{sec:comparison}
In this section, we analyze the performance of NADD against previous adversarial training (AT) and adversarial purification (AP) methods by evaluating their robustness in $\ell_\infty$ and $\ell_2$ norms.
Following this, we analyze the inference times of existing adversarial purification methods. \\

\noindent \textbf{CIFAR-10.}
The left sub-table in Table~\ref{tbl:combined-cifar10} presents the robustness performance against the $\ell_\infty$ norm $(\varepsilon=8/255)$ using PGD+EOT and AutoAttack on CIFAR-10.
The results show that our method outperforms all other diffusion-based purification methods on standard, PGD and AutoAttack robustness accuracies.
In particular, compared to the previous best purification method \cite{lee2023robust}, our method improves robustness accuracy by 3.16\% on WideResNet-28-10, and by 1.79\% on WideResNet-70-16.

Although our technique remains behind adversarial training methods in PGD robust accuracy, it outperforms in  AutoAttack benchmarks.
It is important to to note that adversarial training approaches are specifically trained for the particular p-norm attack used in evaluation, while purification techniques remains independent of the perturbation norm.

The right sub-table in Table~\ref{tbl:combined-cifar10} shows the robustness performance against the $\ell_2$ norm $(\varepsilon=0.5)$ using both PGD+EOT and AutoAttack on CIFAR-10. 
Our method has substantial improvements in both standard and PGD accuracies compared to other adversarial purification methods. 
Additionally, it performs competitively with adversarial training methods, particularly for WRN-28-10, in both standard and PGD robustness accuracies. 
Against AutoAttack, our technique maintains competitive with existing adversarial purification methods.

An additional strength of our technique is its practical applicability. 
As noted, our method achieves this high level of robustness with significantly lower computational overhead, requiring only a single purification diffusion cycle. 
This contrasts with other methods, such as that of Lee \textit{et al.}, which rely on eight cycles to achieve competitive robustness. 
As Table~\ref{tbl:diffcompare} shows, this improvement leads to a $47\times$  reduction in inference time when using a single H100 GPU. \\

Table~\ref{tbl:bpda} shows that our method NADD achieves the highest robustness accuracy against BPDA+EOT $\ell_{\infty}$ norm $(\varepsilon = 8/255)$ attacks on CIFAR-10 with WideResNet-28-10, surpassing Lee et al. \cite{lee2023robust} with one diffusion cycle by nearly 3\% while maintaining competitive standard accuracy. 
This improvement over other methods highlights NADD's effectiveness in providing high resilience to a variety of adversarial attacks.

\noindent \textbf{ImageNet.}
Table~\ref{tbl:inresult} shows that our method achieves a new record in both standard and AutoAttack robustness accuracy, outperforming both adversarial purification techniques and adversarial training  methods. 
While adversarial training methods show competitive robustness, they generally fall short in standard accuracy compared to adversarial purification approaches. 
Among adversarial purification methods, our method consistently yields higher accuracies in both standard and robustness accuracy measures. We conclude that our framework NADD can preserve benign image classification performance, as well as provide enhanced defense against adversarial attacks. 

\begin{table}[t]
\centering
\caption{Standard and robust accuracy against AutoAttack using $\ell_{\infty}$ norm ($\varepsilon=4/255$) on ImageNet. $^\ddagger$Eight diffusion cycles used.}
\small  
\begin{tabular}{llcc}
\toprule
\multirow{2}{*}{Type} & \multirow{2}{*}{Method} & \multicolumn{2}{c}{Accuracy (\%)} \\
\cmidrule(lr){3-4}
& & Standard & Robust \\
\midrule
\multirow{4}{*}{AT} 
  & Salman et al. \cite{salman2020adversarially}& 63.86 & 39.11 \\
  & Bai et al. \cite{bai2021transformers} & 67.38 & 35.51 \\
  & Engstrom et al. \cite{engstrom2019robustness} & 62.42 & 33.20 \\
  & Wong et al. \cite{wongfast2020} & 53.83 & 28.04 \\
\midrule
\multirow{3}{*}{AP}
  & Nie et al. \cite{nie2022diffusion} & 71.48$\pm$0.66 & 38.71$\pm$0.96 \\
  & Lee et al. \cite{lee2023robust}$^\ddagger$ & 70.74$\pm$0.91 & 42.15$\pm$0.64 \\
  & Ours & \textbf{76.23$\pm$0.75} & \textbf{44.23$\pm$0.67}\\
\bottomrule
\end{tabular}
\label{tbl:inresult}
\end{table}


\begin{figure*}[h!]
    \centering
    \subfloat[\footnotesize Impact of High Forward Noise\label{fig3a:ablation}]{\includegraphics[width=0.48\textwidth]{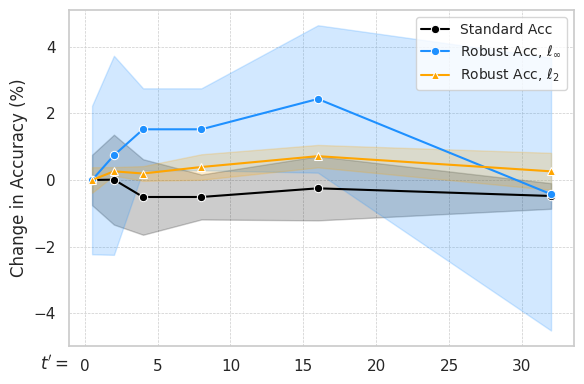}}\hspace{0.01\textwidth}
    \subfloat[\footnotesize Impact of Stopping Correction \label{fig3b:ablation}]{\includegraphics[width=0.48\textwidth]{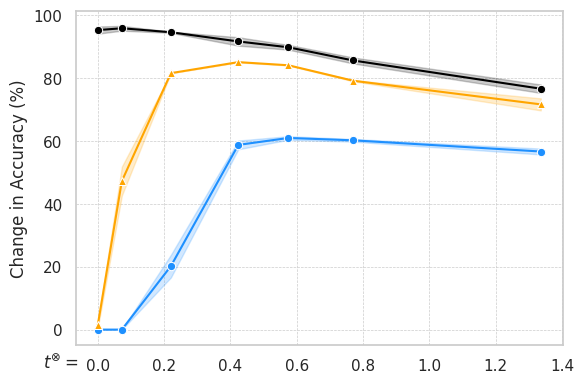}}\\
    \subfloat[\footnotesize Impact of Stochastic Sampling\label{fig3c:ablation}]{\includegraphics[width=0.48\textwidth]{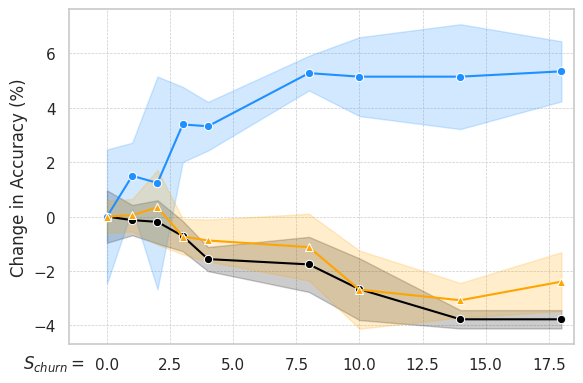}}\hspace{0.01\textwidth}
    \subfloat[\footnotesize Impact of Ring Proximity Condition\label{fig3d:ablation}]{\includegraphics[width=0.48\textwidth]{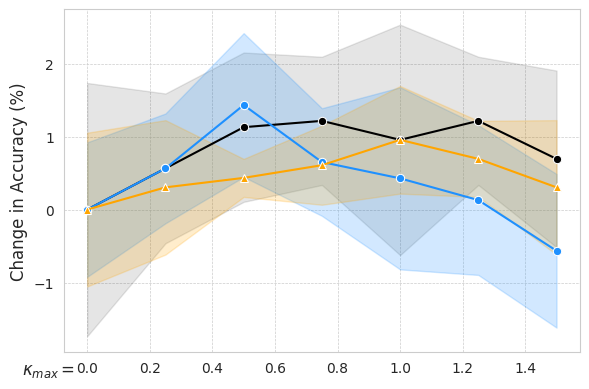}}

    \caption[]{The influence of proposed techniques on standard and robust accuracies against PGD+EOT $\ell_\infty$ ($\varepsilon$=8/255), and $\ell_2$ ($\varepsilon$=0.5), using CIFAR-10 and WideResNet-28-10. Each subplot illustrates the relative change in accuracy (y-axis) based on various factors with 95\% confidence interval. The initial x-axis value serves as the reference point for comparison. \subref{fig3a:ablation} Higher levels of forward noise, as indicated by $t^\prime$, improves robust accuracy against $\ell_\infty$ and $\ell_2$ attacks, while standard accuracy remains mostly unchanged. \subref{fig3b:ablation} Stopping correction within the range $t^\otimes \in [0.4,0.6]$ significantly enhances robust accuracy against both $\ell_\infty$ and $\ell_2$ norms. \subref{fig3c:ablation} Increasing reverse noise improves robustness against $\ell_\infty$ attacks however can reduce robustness against $\ell_2$ attacks if too high. \subref{fig3d:ablation} This sub-figure shows the robustness of a NADD model evaluated using a radius of $\kappa_{max}-\kappa_{min} = 0.25$. Increasing the ring radius improves robust accuracy however begins to decrease after $\kappa_{max}=1.0$.}
    \label{fig3:ablation}
\end{figure*}



\subsection{Ablation Study}
\label{sec:ablation}

This section presents an investigation into the effect of our proposed techniques on robustness accuracy of WideResNet28-10 against PGD+EOT $\ell_\infty$ and $\ell_2$ norms using CIFAR10. 
Unless otherwise specified, the forward diffusion time-step is set to $t^\prime=16$, the time-step stop correction to $t^\otimes=0.585$ and stochastic sampling noise to $S_{churn}=0$.
The main results of this ablation study are presented in Figure~\ref{fig3:ablation} with an analysis and discussion below.
\newline 

\noindent \textbf{Impact of High Forward Noise ($t^\prime$).} 
In this experiment, we evaluate different levels of forward noise by varying $t^\prime \in [0.5, 32]$.
We observe three patterns from the results presented in Figure~\ref{fig3a:ablation}.
First, we find that increasing noise in the forward diffusion process improves robustness and then begins to decrease.
The ideal amount of noise for both norm settings is $t^\prime=16$.
Second, the improvement in robustness is more pronounced against $\ell_\infty$ attacks than $\ell_2$ attacks.
And lastly, standard accuracy is not significantly impacted by increasing forward noise.
This demonstrates that we are able to improve the noise-reconstruction trade-off due to the introduction of the scheduled correction update mechanism.
\newline 

\noindent \textbf{Impact of Stopping Correction ($t^\otimes$).}
Here, we evaluate different time-steps for stopping the correction updates between $t^\otimes \in [0.0,1.5]$. 
The results in  Figure~\ref{fig3b:ablation} also reveal three patterns. 
First, it shows that without a stopping correction, where $t^\otimes=0$, the robustness accuracy is close to zero because the diffusion model is reconstructing the adversarial perturbation. 
Second, the robustness accuracy against both norm settings improves as $t^\otimes$ increases, and then begins to decrease for only the $\ell_2$ norm at $t^\otimes=0.434$.
Importantly, the standard accuracy decreases as $t^\otimes$ increases, revealing a key trade-off. 
Considering this factor, the optimal time-step for stopping correction updates should keep standard accuracy above 90\% and differs between attack models ($\ell_2$: $t^\otimes=0.434$, $\ell_\infty$: $t^\otimes=0.585$).
\newline 

\noindent \textbf{Impact of Stochastic Sampling ($S_{churn}$).}
We consider the role of reverse process stochasticity against adversarial attacks by evaluating different values of $S_{churn} \in [0, 18]$.
As seen in Figure~\ref{fig3c:ablation}, increasing  $S_{churn}$ results in significantly higher robustness accuracy against $\ell_\infty$ attacks. 
The improvements in robustness against $\ell_2$ attacks are less pronounced and begin to decrease at low values of $S_{churn}$. 
We also observe that standard accuracy decreases as the level of stochasticity increases which reveals another trade-off. 
Thus, the ideal level of stochasticity during sampling also varies here between attack models ($\ell_2$: $S_{churn}=8$, $\ell_\infty$: $S_{churn}=2$). \\

\noindent \textbf{Impact of Ring Proximity Radius ($\kappa_{min}, \kappa_{max}$).}
In this experiment, we analyze the effect of varying the ring proximity radius, specifically by adjusting the parameters $\kappa_{min}$ and $\kappa_{max}$, to observe the changes in robustness accuracy against both $\ell_\infty$ and $\ell_2$ norms. Figure~\ref{fig3d:ablation} presents the results, which shows that increasing $\kappa_{max}$ initially improves robustness accuracy across both norms settings but eventually leads to a decline, indicating an optimal radius exists. 
This optimal range lies between $\kappa_{max}=(0.4, 1.0)$. 
The improvements are more pronounced against the $\ell_\infty$ norm compared to the $\ell_2$ norm, suggesting that the proximity radius contributes differently to robustness based on the norm setting. Beyond $\kappa_{max}=1.0$, standard accuracy begins to deteriorate, revealing a critical trade-off where larger proximity radii compromise model reliability. Therefore, an ideal setting for $\kappa_{max}$ may be approximately 0.5 to 0.6 for $\ell_\infty$ robustness while maintaining reasonable accuracy against $\ell_2$ norm.

\section{Discussion}
The results of our study illustrate a significant advancement in the field of adversarial purification.
By amplifying noise levels guided with new techniques, such as stochastic sampling and the ring proximity condition, our NADD framework achieves superior performance against sophisticated adversarial attacks. 
In this section, we reflect on the implications, and potential future directions of this work. \\

\noindent \textbf{Practical Implications.} The improvements in robust accuracy underscore the practicality of employing NADD in real-world security-sensitive domains, such as autonomous vehicles and medical imaging systems. 
Unlike traditional adversarial training methods, which are tailored for specific attack types and require extensive computational resources for retraining, our approach is model-agnostic and adapts to different attack settings without significant overhead.
The reduced inference time, facilitated by fewer diffusion steps, positions NADD as a viable solution for real-time applications where speed and reliability are paramount.
Future work could explore lightweight versions of NADD, potentially leveraging model compression techniques or hybrid architectures that strike a balance between performance and computational load. \\

\noindent \textbf{Comparison with Existing Purification Approaches.}
Existing state-of-the-art purification methods, such as DiffPure and GDMP, rely on limited noise levels during the forward diffusion process to minimize reconstruction error and preserve input semantics. 
However, these conservative approaches fall short when facing more stronger adversarial attacks. 
Our work bridges this gap by demonstrating that controlled amplification of noise, paired with targeted corrective strategies, can effectively counter such vulnerabilities. 
This robustness enhancement comes with a trade-off in standard accuracy, which remains within acceptable bounds, ensuring that the purified output remains semantically similar to the original input.\\


\noindent \textbf{Future Directions.}
While NADD has shown resilience to white-box attacks, the landscape of adversarial strategies is rapidly evolving. 
Adaptive attackers, capable of leveraging insights into our purification strategy, may attempt to tailor perturbations that exploit potential weaknesses in noise amplification or correction mechanisms. 
Further research should investigate the resilience of NADD against such adaptive adversaries and explore adaptive learning mechanisms that allow the model to update its purification strategy based on the evolving threat landscape.
Additionally, future work could delve into integrating NADD with one-shot diffusion models, such as Consistency models\cite{song2023consistency}, or other complementary defense mechanisms, such as randomized smoothing or ensemble approaches, to create a layered defense that maximizes robustness while maintaining efficiency.

\section{Conclusion}

In this work, we introduced a novel adversarial purification framework, Noise Amplified Diffusion Defence (NADD), which systematically enhances the robustness of classifiers against adversarial attacks by incorporating higher levels of noise during both the forward and reverse diffusion processes. 
By leveraging the ring proximity condition, we improved the trade-off between reconstruction quality and adversarial robustness. 
Our approach introduces stochastic sampling and correction schedules to preserve semantic features while effectively eliminating adversarial noise. 
Benchmarking results demonstrated that NADD surpasses existing purification methods in terms of both robust accuracy and computational efficiency, achieving significant reductions in inference time. 
These findings underline the potential of diffusion-based defenses to strengthen neural network robustness in practical, safety-critical applications without resorting to gradient obfuscation techniques. 
Future research could explore the integration of adaptive noise levels and alternative diffusion frameworks to further enhance the scalability and performance of adversarial purification strategies.
\bibliographystyle{IEEEtran}
%



\bibliography{reference}

@book{abramowitz-handbook,
author = {Abramowitz, Milton},
title = {Handbook of Mathematical Functions, With Formulas, Graphs, and Mathematical Tables,},
year = {1974},
isbn = {0486612724},
publisher = {Dover Publications, Inc.},
address = {USA}
}

@inproceedings{xiao2023densepure,
  title={Densepure: Understanding diffusion models for adversarial robustness},
  author={Xiao, Chaowei and Chen, Zhongzhu and Jin, Kun and Wang, Jiongxiao and Nie, Weili and Liu, Mingyan and Anandkumar, Anima and Li, Bo and Song, Dawn},
  booktitle={The Eleventh International Conference on Learning Representations},
  year={2023}
}

@article{karras2022elucidating,
  title={Elucidating the design space of diffusion-based generative models},
  author={Karras, Tero and Aittala, Miika and Aila, Timo and Laine, Samuli},
  journal={Advances in Neural Information Processing Systems},
  volume={35},
  pages={26565--26577},
  year={2022}
}

@inproceedings{
  song2020score,
  title={Score-Based Generative Modeling through Stochastic Differential Equations},
  author={Yang Song and Jascha Sohl-Dickstein and Diederik P Kingma and Abhishek Kumar and Stefano Ermon and Ben Poole},
  booktitle={International Conference on Learning Representations (ICLR)},
  year={2021},
  url={https://openreview.net/forum?id=PxTIG12RRHS}
}

@article{ho2020denoising,
  title={Denoising diffusion probabilistic models},
  author={Ho, Jonathan and Jain, Ajay and Abbeel, Pieter},
  journal={Advances in neural information processing systems},
  volume={33},
  pages={6840--6851},
  year={2020}
}

@inproceedings{sohl2015deep,
  title={Deep unsupervised learning using nonequilibrium thermodynamics},
  author={Sohl-Dickstein, Jascha and Weiss, Eric and Maheswaranathan, Niru and Ganguli, Surya},
  booktitle={International conference on machine learning},
  pages={2256--2265},
  year={2015},
  organization={PMLR}
}

@inproceedings{carlini2022certified,
  title={(Certified!!) Adversarial Robustness for Free!},
  author={Carlini, Nicholas and Tramer, Florian and Dvijotham, Krishnamurthy Dj and Rice, Leslie and Sun, Mingjie and Kolter, J Zico},
  booktitle={The Eleventh International Conference on Learning Representations},
year={2022}
}

@inproceedings{nie2022diffusion,
  title={Diffusion Models for Adversarial Purification},
  author={Nie, Weili and Guo, Brandon and Huang, Yujia and Xiao, Chaowei and Vahdat, Arash and Anandkumar, Animashree},
  booktitle={International Conference on Machine Learning},
  pages={16805--16827},
  year={2022},
  organization={PMLR}
}

@article{hyvarinen2005estimation,
  title={Estimation of non-normalized statistical models by score matching.},
  author={Hyv{\"a}rinen, Aapo and Dayan, Peter},
  journal={Journal of Machine Learning Research},
  volume={6},
  number={4},
  year={2005}
}

@inproceedings{song2018pixeldefend,
  title={PixelDefend: Leveraging Generative Models to Understand and Defend against Adversarial Examples},
  author={Song, Yang and Kim, Taesup and Nowozin, Sebastian and Ermon, Stefano and Kushman, Nate},
  booktitle={International Conference on Learning Representations},
  year={2018}
}

@inproceedings{yoon2021adversarial,
  title={Adversarial purification with score-based generative models},
  author={Yoon, Jongmin and Hwang, Sung Ju and Lee, Juho},
  booktitle={International Conference on Machine Learning},
  pages={12062--12072},
  year={2021},
  organization={PMLR}
}

@article{song2020improved,
  title={Improved techniques for training score-based generative models},
  author={Song, Yang and Ermon, Stefano},
  journal={Advances in neural information processing systems},
  volume={33},
  pages={12438--12448},
  year={2020}
}

@inproceedings{song2023consistency,
  title={Consistency Models},
  author={Song, Yang and Dhariwal, Prafulla and Chen, Mark and Sutskever, Ilya},
  booktitle={International Conference on Machine Learning},
  pages={32211--32252},
  year={2023},
  organization={PMLR}
}

@inproceedings{dockhorn2021score,
  title={Score-Based Generative Modeling with Critically-Damped Langevin Diffusion},
  author={Dockhorn, Tim and Vahdat, Arash and Kreis, Karsten},
  booktitle={International Conference on Learning Representations},
  year={2021}
}

@inproceedings{paschali2018generalizability,
  title={Generalizability vs. robustness: investigating medical imaging networks using adversarial examples},
  author={Paschali, Magdalini and Conjeti, Sailesh and Navarro, Fernando and Navab, Nassir},
  booktitle={Medical Image Computing and Computer Assisted Intervention--MICCAI 2018: 21st International Conference, Granada, Spain, September 16-20, 2018, Proceedings, Part I},
  pages={493--501},
  year={2018},
  organization={Springer}
}

@article{kaviani2022adversarial,
  title={Adversarial attacks and defenses on AI in medical imaging informatics: A survey},
  author={Kaviani, Sara and Han, Ki Jin and Sohn, Insoo},
  journal={Expert Systems with Applications},
  volume={198},
  pages={116815},
  year={2022},
  publisher={Elsevier}
}

@inproceedings{zhang2022adversarial,
  title={On adversarial robustness of trajectory prediction for autonomous vehicles},
  author={Zhang, Qingzhao and Hu, Shengtuo and Sun, Jiachen and Chen, Qi Alfred and Mao, Z Morley},
  booktitle={Proceedings of the IEEE/CVF Conference on Computer Vision and Pattern Recognition},
  pages={15159--15168},
  year={2022}
}

@article{michaelis2019benchmarking,
  title={Benchmarking robustness in object detection: Autonomous driving when winter is coming},
  author={Michaelis, Claudio and Mitzkus, Benjamin and Geirhos, Robert and Rusak, Evgenia and Bringmann, Oliver and Ecker, Alexander S and Bethge, Matthias and Brendel, Wieland},
  journal={arXiv preprint arXiv:1907.07484},
  year={2019}
}

@article{salman2020denoised,
  title={Denoised smoothing: A provable defense for pretrained classifiers},
  author={Salman, Hadi and Sun, Mingjie and Yang, Greg and Kapoor, Ashish and Kolter, J Zico},
  journal={Advances in Neural Information Processing Systems},
  volume={33},
  pages={21945--21957},
  year={2020}
}

@article{szegedy2013intriguing,
  title={Intriguing properties of neural networks},
  author={Szegedy, Christian and Zaremba, Wojciech and Sutskever, Ilya and Bruna, Joan and Erhan, Dumitru and Goodfellow, Ian and Fergus, Rob},
  journal={arXiv preprint arXiv:1312.6199},
  year={2013}
}

@article{goodfellow2014explaining,
  title={Explaining and harnessing adversarial examples},
  author={Goodfellow, Ian J and Shlens, Jonathon and Szegedy, Christian},
  journal={arXiv preprint arXiv:1412.6572},
  year={2014}
}

@article{madry2017towards,
  title={Towards deep learning models resistant to adversarial attacks},
  author={Madry, Aleksander and Makelov, Aleksandar and Schmidt, Ludwig and Tsipras, Dimitris and Vladu, Adrian},
  journal={arXiv preprint arXiv:1706.06083},
  year={2017}
}

@article{krizhevsky2009learning,
  title={Learning multiple layers of features from tiny images},
  author={Krizhevsky, Alex and Hinton, Geoffrey and others},
  year={2009},
  publisher={Toronto, ON, Canada}
}

@inproceedings{croce2022evaluating,
  title={Evaluating the adversarial robustness of adaptive test-time defenses},
  author={Croce, Francesco and Gowal, Sven and Brunner, Thomas and Shelhamer, Evan and Hein, Matthias and Cemgil, Taylan},
  booktitle={International Conference on Machine Learning},
  pages={4421--4435},
  year={2022},
  organization={PMLR}
}

@article{zagoruyko2016wide,
  title={Wide residual networks},
  author={Zagoruyko, Sergey and Komodakis, Nikos},
  journal={arXiv preprint arXiv:1605.07146},
  year={2016}
}

@inproceedings{he2016deep,
  title={Deep residual learning for image recognition},
  author={He, Kaiming and Zhang, Xiangyu and Ren, Shaoqing and Sun, Jian},
  booktitle={Proceedings of the IEEE conference on computer vision and pattern recognition},
  pages={770--778},
  year={2016}
}

@inproceedings{deng2009imagenet,
  title={Imagenet: A large-scale hierarchical image database},
  author={Deng, Jia and Dong, Wei and Socher, Richard and Li, Li-Jia and Li, Kai and Fei-Fei, Li},
  booktitle={2009 IEEE conference on computer vision and pattern recognition},
  pages={248--255},
  year={2009},
  organization={Ieee}
}

@inproceedings{athalye2018obfuscated,
  title={Obfuscated gradients give a false sense of security: Circumventing defenses to adversarial examples},
  author={Athalye, Anish and Carlini, Nicholas and Wagner, David},
  booktitle={International conference on machine learning},
  pages={274--283},
  year={2018},
  organization={PMLR}
}

@inproceedings{hillstochastic,
  title={Stochastic Security: Adversarial Defense Using Long-Run Dynamics of Energy-Based Models},
  author={Hill, Mitch and Mitchell, Jonathan Craig and Zhu, Song-Chun},
  booktitle={International Conference on Learning Representations},
  year={2020}
}

@article{wang2022guided,
  title={Guided diffusion model for adversarial purification},
  author={Wang, Jinyi and Lyu, Zhaoyang and Lin, Dahua and Dai, Bo and Fu, Hongfei},
  journal={arXiv preprint arXiv:2205.14969},
  year={2022}
}

@inproceedings{lee2023robust,
  title={Robust evaluation of diffusion-based adversarial purification},
  author={Lee, Minjong and Kim, Dongwoo},
  booktitle={Proceedings of the IEEE/CVF International Conference on Computer Vision},
  pages={134--144},
  year={2023}
}

@article{srinivasan2021robustifying,
  title={Robustifying models against adversarial attacks by langevin dynamics},
  author={Srinivasan, Vignesh and Rohrer, Csaba and Marban, Arturo and M{\"u}ller, Klaus-Robert and Samek, Wojciech and Nakajima, Shinichi},
  journal={Neural Networks},
  volume={137},
  pages={1--17},
  year={2021},
  publisher={Elsevier}
}

@article{yang2019me,
  title={Me-net: Towards effective adversarial robustness with matrix estimation},
  author={Yang, Yuzhe and Zhang, Guo and Katabi, Dina and Xu, Zhi},
  journal={arXiv preprint arXiv:1905.11971},
  year={2019}
}

@inproceedings{athalye2018synthesizing,
  title={Synthesizing robust adversarial examples},
  author={Athalye, Anish and Engstrom, Logan and Ilyas, Andrew and Kwok, Kevin},
  booktitle={International conference on machine learning},
  pages={284--293},
  year={2018},
  organization={PMLR}
}

@article{dhariwal2021diffusion,
  title={Diffusion models beat gans on image synthesis},
  author={Dhariwal, Prafulla and Nichol, Alexander},
  journal={Advances in neural information processing systems},
  volume={34},
  pages={8780--8794},
  year={2021}
}

@article{gowal2020uncovering,
  title={Uncovering the limits of adversarial training against norm-bounded adversarial examples},
  author={Gowal, Sven and Qin, Chongli and Uesato, Jonathan and Mann, Timothy and Kohli, Pushmeet},
  journal={arXiv preprint arXiv:2010.03593},
  year={2020}
}

@article{gowal2021improving,
  title={Improving robustness using generated data},
  author={Gowal, Sven and Rebuffi, Sylvestre-Alvise and Wiles, Olivia and Stimberg, Florian and Calian, Dan Andrei and Mann, Timothy A},
  journal={Advances in Neural Information Processing Systems},
  volume={34},
  pages={4218--4233},
  year={2021}
}

@article{rebuffi2021fixing,
  title={Fixing data augmentation to improve adversarial robustness},
  author={Rebuffi, Sylvestre-Alvise and Gowal, Sven and Calian, Dan A and Stimberg, Florian and Wiles, Olivia and Mann, Timothy},
  journal={arXiv preprint arXiv:2103.01946},
  year={2021}
}

@inproceedings{pang2022robustness,
  title={Robustness and accuracy could be reconcilable by (proper) definition},
  author={Pang, Tianyu and Lin, Min and Yang, Xiao and Zhu, Jun and Yan, Shuicheng},
  booktitle={International Conference on Machine Learning},
  pages={17258--17277},
  year={2022},
  organization={PMLR}
}

@inproceedings{augustin2020adversarial,
  title={Adversarial robustness on in-and out-distribution improves explainability},
  author={Augustin, Maximilian and Meinke, Alexander and Hein, Matthias},
  booktitle={European Conference on Computer Vision},
  pages={228--245},
  year={2020},
  organization={Springer}
}

@article{engstrom2019robustness,
  title={Robustness (python library), 2019},
  author={Engstrom, Logan and Ilyas, Andrew and Salman, Hadi and Santurkar, Shibani and Tsipras, Dimitris},
  journal={URL https://github. com/MadryLab/robustness},
  volume={4},
  number={4},
  pages={4--3},
  year={2019}
}

@article{salman2020adversarially,
  title={Do adversarially robust imagenet models transfer better?},
  author={Salman, Hadi and Ilyas, Andrew and Engstrom, Logan and Kapoor, Ashish and Madry, Aleksander},
  journal={Advances in Neural Information Processing Systems},
  volume={33},
  pages={3533--3545},
  year={2020}
}

@article{bai2021transformers,
  title={Are transformers more robust than cnns?},
  author={Bai, Yutong and Mei, Jieru and Yuille, Alan L and Xie, Cihang},
  journal={Advances in neural information processing systems},
  volume={34},
  pages={26831--26843},
  year={2021}
}

@inproceedings{wongfast2020,
  title={Fast is better than free: Revisiting adversarial training},
  author={Wong, Eric and Rice, Leslie and Kolter, J Zico},
  booktitle={International Conference on Learning Representations},
  year={2020}
}

@article{hill2020stochastic,
  title={Stochastic security: Adversarial defense using long-run dynamics of energy-based models},
  author={Hill, Mitch and Mitchell, Jonathan and Zhu, Song-Chun},
  journal={arXiv preprint arXiv:2005.13525},
  year={2020}
}

@inproceedings{karras2024analyzing,
  title={Analyzing and improving the training dynamics of diffusion models},
  author={Karras, Tero and Aittala, Miika and Lehtinen, Jaakko and Hellsten, Janne and Aila, Timo and Laine, Samuli},
  booktitle={Proceedings of the IEEE/CVF Conference on Computer Vision and Pattern Recognition},
  pages={24174--24184},
  year={2024}
}

@inproceedings{tramer2017ensemble,
  title={Ensemble Adversarial Training: Attacks and Defenses},
  author={Tram{\`e}r, Florian and Kurakin, Alexey and Papernot, Nicolas and Goodfellow, Ian and Boneh, Dan and McDaniel, Patrick},
  booktitle={International Conference on Learning Representations},
  year={2018}
}

@article{volpi2018generalizing,
  title={Generalizing to unseen domains via adversarial data augmentation},
  author={Volpi, Riccardo and Namkoong, Hongseok and Sener, Ozan and Duchi, John C and Murino, Vittorio and Savarese, Silvio},
  journal={Advances in neural information processing systems},
  volume={31},
  year={2018}
}

@article{rebuffi2021data,
  title={Data augmentation can improve robustness},
  author={Rebuffi, Sylvestre-Alvise and Gowal, Sven and Calian, Dan Andrei and Stimberg, Florian and Wiles, Olivia and Mann, Timothy A},
  journal={Advances in Neural Information Processing Systems},
  volume={34},
  pages={29935--29948},
  year={2021}
}

@article{hendrycks2021unsolved,
  title={Unsolved problems in ml safety},
  author={Hendrycks, Dan and Carlini, Nicholas and Schulman, John and Steinhardt, Jacob},
  journal={arXiv preprint arXiv:2109.13916},
  year={2021}
}

@article{bai2021recent,
  title={Recent advances in adversarial training for adversarial robustness},
  author={Bai, Tao and Luo, Jinqi and Zhao, Jun and Wen, Bihan and Wang, Qian},
  journal={arXiv preprint arXiv:2102.01356},
  year={2021}
}

@article{xie2019intriguing,
  title={Intriguing properties of adversarial training at scale},
  author={Xie, Cihang and Yuille, Alan},
  journal={arXiv preprint arXiv:1906.03787},
  year={2019}
}

@article{samangouei2018defense,
  title={Defense-gan: protecting classifiers against adversarial attacks using generative models},
  author={Samangouei, P},
  journal={arXiv preprint arXiv:1805.06605},
  year={2018}
}

@article{lin2024robust,
  title={Robust Diffusion Models for Adversarial Purification},
  author={Lin, Guang and Tao, Zerui and Zhang, Jianhai and Tanaka, Toshihisa and Zhao, Qibin},
  journal={arXiv preprint arXiv:2403.16067},
  year={2024}
}

@article{zhang2024classifier,
  title={Classifier Guidance Enhances Diffusion-based Adversarial Purification by Preserving Predictive Information},
  author={Zhang, Mingkun and Li, Jianing and Chen, Wei and Guo, Jiafeng and Cheng, Xueqi},
  journal={arXiv preprint arXiv:2408.05900},
  year={2024}
}

@article{sehwag2021robust,
  title={Robust learning meets generative models: Can proxy distributions improve adversarial robustness?},
  author={Sehwag, Vikash and Mahloujifar, Saeed and Handina, Tinashe and Dai, Sihui and Xiang, Chong and Chiang, Mung and Mittal, Prateek},
  journal={arXiv preprint arXiv:2104.09425},
  year={2021}
}

\newpage 
\begin{appendices}
\section{Theoretical Justification}
\label{appx:proofs}

We give details for the proof of Theorem~ \ref{thm:estimates}.
    In the forward diffusion process ${\bf x} = {\bf x}_{t_0} \to $ ${\bf x}_{t_1} \to $ $\ldots$ $ \to {\bf x}_{t_N}$, each update step ${\bf x}_{t_i} \to $ ${\bf x}_{t_{i+1}}$ is an addition with a Gaussian vector: ${\bf x}_{t_{i+1}} = {\bf x}_{t_{i}} + {\bf z}_i$, where ${\bf z}_i \sim \mathcal{N}( {\bf 0}, {\bf I}(t_{i+1}^2 - t_i^2 ))$. 
    In the reverse process with correction ${\bf x}_{t_0} \leftarrow $ ${\bf \hat{x}}_{t_1} \leftarrow$ $\ldots$ $ \leftarrow {\bf \hat{x}}_{t_N} = {\bf x}_{t_N}$, the update  ${\bf x}_{t_i} \leftarrow $ ${\bf \hat{x}}_{t_{i+1}}$ is given in Eq. (\ref{eq:correction}).

    The intuition in our proof is the following. First, we derive a formula of ${\bf x}_{t_i} - {\bf \hat{x}}_{t_i} $ as a scaled sum of $ {\bf x}_{t_{i+1}} - {\bf \hat{x}}_{t_{i+1}}$ with Gaussian vectors. Second, due to the concentration of Gaussian random vectors, with high probability, the difference between ${\bf x}_{t_i} - {\bf \hat{x}}_{t_i} $ and $ {\bf x}_{t_{i+1}} - {\bf \hat{x}}_{t_{i+1}}$ is small.
    By applying union bound and concentration estimates of Gaussian random variables, we can obtain a lower bound for the probability ${\sf Pr}[ \|{\bf x} - {\bf x}_{t^\otimes	} \|_2 \leq \kappa_{\text{max}} ]$, and fine-tuning $w_{i}$'s will give the desired estimate.
    Finally, to lower bound ${\sf Pr}[ \|{\bf x} - {\bf x}_{t^\otimes	} \|_2 \leq \kappa_{\text{min}} ]$, we fine-tune with a different choice of $w_{i}$'s so that the Gaussian probability in the hypercube centered at the origin, of length $\kappa_{\text{min}}$ is less than $1- \delta_*$.
    For simplicity, we assume that $t^\otimes	 = t_0 = 0$, $t' = t_N = T$, and argue for the general case $t^\otimes	 < \varepsilon$ and $t' < T$ at the end.

    Using the defining formulas of ${\bf x}_{t_i}$, ${\bf \hat{x}}_{t_i}$ in Eq.~(\ref{eq:forwar-step}) and Eq.~(\ref{eq:correction}), we expand $ {\bf x}_{t_i} - {\bf \hat{x}}_{t_i} $ as:
    {\scriptsize
    \begin{align}
        & {\bf x}_{t_i} - {\bf \hat{x}}_{t_i} 
        = 
        - \big( {\bf x}_{t_{i+1}} - {\bf z}_i \big)
        \notag 
        \\
        &
        -
        \big( {\bf \hat{x}}_{t_{i+1}} + (t_i - t_{i+1}) \big[\Phi(\mathbf{x}_{t_{i+1}} , t_{i+1} ; \theta)(1-w_i)  + c_{i}w_i \big]\big) 
        \notag \\
        & = 
        \big[{\bf x}_{t_{i+1}} - {\bf \hat{x}}_{t_{i+1}} - {\bf z}_i - (t_i - t_{i+1}) \Phi(\hat{\mathbf{x}}_{t_{i+1}}' , t_{i+1} ; \theta) \big](1-w_i) \label{eq:thm2}
    \end{align}
    }
    where the correction schedule weight $w_i$ is defined in Eq.~(15), and the stochastic sampling $\hat{\mathbf{x}}_{t_{i+1}}'$ is given in Eq.~(17).

    Next, by writing the update $\Phi$ in terms of the denoiser $D_{\theta}$ for Euler solver, we can transform $(t_i - t_{i+1}) \Phi(\hat{\mathbf{x}}_{t_{i+1}}', t_{i+1} ; \theta)$ as follows:
    \begin{align*}
        & (t_i - t_{i+1}) \Phi(\hat{\mathbf{x}}_{t_{i+1}}', t_{i+1} ; \theta) \\
      = & (t_i - t_{i+1}) \big[ \mathbf{\hat{x}}_{t_{i+1}}' -  D_\theta(\mathbf{\hat{x}}_{t_{i+1}}';{t_{i+1}}) \big]/t_{i+1} \\
      = & (\frac{t_i}{t_{i+1}} - 1) \big[ \mathbf{\hat{x}}_{t_{i+1}}' -  (\nabla_{ {\bf x} } \log p_{t_{i+1}}( {\bf \hat{x}}_{t_{i+1}}') t_{i+1}^2 + {\bf \hat{x}}_{t_{i+1}})' \big] \\
      = & - (t_i - t_{i+1}) t_{i+1} \nabla_{ {\bf x} } \log p_{t_{i+1}}( {\bf \hat{x}}_{t_{i+1}}') 
    \end{align*}

    The probability flow ODE \cite{song2020score} from Eq. (1) in \cite{karras2022elucidating} with $\sigma(t) = t$ and $s(t) = 1$ is given by 
    $
        t \cdot \nabla_{ {\bf x} } \log p_{t}( {\bf \hat{x}}_{t}) = - \frac{ {\sf d} {\bf \hat{x}}_t }{ {\sf d} t }
    $,
    which is discretized by the Euler solver in our update function as 
    \begin{equation*}
        t_{i+1} \cdot \nabla_{ {\bf x} } \log p_{t_{i+1}}( {\bf \hat{x}}_{t_{i+1}}') = - \frac{ {\bf \hat{x}}_{t_{i+1}}' -  {\bf \hat{x}}_{t_{i}}' }{ t_{i+1} - t_i }
    \end{equation*}
    or equivalently 
     \begin{equation}
          - (t_i - t_{i+1}) t_{i+1} \cdot \nabla_{ {\bf x} } \log p_{t_{i+1}}( {\bf \hat{x}}_{t_{i+1}}) = {\bf \hat{x}}_{t_{i}}' - {\bf \hat{x}}_{t_{i+1}}'
     \end{equation}

     Due to the stochastic sampling component, we treat to three separated cases

     \begin{enumerate}
         \item If $t_i, t_{i-1} \in [S_{ \text{min} }, S_{ \text{max} }]$, then ${\bf \hat{x}}_{t_{i}}' - {\bf \hat{x}}_{t_{i+1}}'$ is approximately a Gaussian random vector ${\bf z}_i' \sim \mathcal{N}(\mathbf{0}, \mathbf{I}( (1+\gamma)^2t_{i+1}^2 - (1+\gamma)^2t_{i}^2))$.
         \item If $t_{i} \in [S_{ \text{min} }, S_{ \text{max} }]$, but $t_{i}-1$ is not, then ${\bf \hat{x}}_{t_{i}}' - {\bf \hat{x}}_{t_{i+1}}'$ is approximately a Gaussian random vector ${\bf z}_i' \sim \mathcal{N}(\mathbf{0}, \mathbf{I}( (1+\gamma)^2t_{i+1}^2 - t_{i}^2))$.
         \item If $t_{i}, t_{i-1} \notin [S_{ \text{min} }, S_{ \text{max} }]$, then ${\bf \hat{x}}_{t_{i}}' - {\bf \hat{x}}_{t_{i+1}}'$ is approximately a Gaussian random vector ${\bf z}_i' \sim \mathcal{N}(\mathbf{0}, \mathbf{I} (t_{i+1}^2 - t_{i}^2))$.
     \end{enumerate}

     Since the stochastic sampling coefficient $\gamma$ is choosen depending only on the adversarial norm, and independent of the denoising procedure, we can prove the third case ($\gamma = 0$), then adjust the weights $w_i$, by increasing $w_i$ in other cases when $\gamma>0$, to obtain the same estimate on ${\sf Pr}[ \|{\bf x} - {\bf x}_{t^\otimes	} \|_2 \leq \kappa_{\text{max}} ]$.
     
     Overall, this gives  $(t_i - t_{i+1}) \Phi(\mathbf{x}_{t_{i+1}} , t_{i+1} ; \theta) = {\bf z}_i' \sim \mathcal{N}(\mathbf{0}, \mathbf{I} (t_{i+1}^2 - t_{i}^2))$ which we substitute in Eq. (\ref{eq:thm2}) to obtain the following equation for ${\bf x}_{t_i} - {\bf \hat{x}}_{t_i}$: 
     \begin{align} \label{eq:thm1}
         {\bf x}_{t_i} - {\bf \hat{x}}_{t_i} 
        =
        \big[({\bf x}_{t_{i+1}} - {\bf \hat{x}}_{t_{i+1}}) - {\bf z}_i - {\bf z}_i' \big](1-w_i) 
    \end{align}
    where $ {\bf z}_i, {\bf z}_i' \sim \mathcal{N}(\mathbf{0}, \mathbf{I} (t_{i+1}^2 - t_{i}^2))$.

    \textit{Upper-bound by $\kappa_{ \text{max} }$:}
     We apply the following simple variant of the union bound for two non-negative random variables $X, X'$ to our case with $X = \|  {\bf x}_{t_i} - {\bf \hat{x}}_{t_i} \|$ and $X' = {\bf z}_i + {\bf z}_i' \sim \mathcal{N}( {\bf 0}, {\bf I} \cdot 2(t_i^2 - t_{i+1}^2) )$: if
    $
    {\sf Pr}[ X \leq \varepsilon ] \geq 1- \delta \text{ and }
    {\sf Pr}[ X' \leq \varepsilon' ] \geq 1 - \delta',
    $
    then 
    \begin{equation}\label{eq:thm3}
         {\sf Pr}[ X+X' \leq \varepsilon + \varepsilon'] \geq 1 - (\delta + \delta')
    \end{equation}
   One can prove the bound in Eq. (\ref{eq:thm3}) via the following equivalent inequality: for any two non-negative random variables $X, X'$ such that 
    $
    {\sf Pr}\Big[ X \geq \varepsilon \Big] \leq \delta \text{ and }
    {\sf Pr} \Big[ X' \geq \varepsilon' \Big] \leq \delta',
    $
    we have
    \begin{equation}
         {\sf Pr} \Big[ X+X' \geq \varepsilon + \varepsilon' \Big] \leq \delta + \delta'
    \end{equation}
    Indeed, if $(X - \varepsilon) + (X' - \varepsilon') \geq 0$, then one must have either $(X - \varepsilon) \geq 0$ or $ (X' - \varepsilon') \geq 0$, otherwise the sum cannot be non-negative. An union bound then gives the desired inequality.

    Next, for induction from step $i+1$ to step $i$, let us assume that we already have an estimate 
    \begin{equation}
        {\sf Pr} \Big[ \| {\bf x}_{t_{i+1}} - {\bf \hat{x}}_{t_{i+1}}\|_2 \leq \varepsilon_{i+1} \Big] \geq 1 - \delta_{i+1}
    \end{equation}
    Our goal in the induction step is to prove 
    \begin{equation}
        {\sf Pr}\Big[ \| {\bf x}_{t_{i}} - {\bf \hat{x}}_{t_{i}}\|_2 \leq \varepsilon_{i} \Big] \geq 1 - \delta_{i}
    \end{equation}
    for some $\varepsilon_i$, $\delta_i$ depends on $\varepsilon_{i+1}$, $\delta_{i+1}$, $t_1, t_{i+1}$ and $w_i$. 

    We now utilize the Eq. (\ref{eq:thm1}) above. By using concentration inequality for Gaussian random vectors, for any $\lambda_{i+1} \geq 0$ which can be chosen later, it holds that 
    \begin{equation*}
        {\sf Pr}\Big[ \| {\bf z}_i + {\bf z}_i' \|_2 \leq \lambda_{i+1}\Big] \geq 1 - 2e^{ - \frac{\lambda_{i+1}^2}{4(-t_i^2 + t_{i+1}^2)} }
    \end{equation*}

    If $ \| {\bf x}_{t_{i+1}} - {\bf \hat{x}}_{t_{i+1}}\|_2 + \| {\bf z}_i + {\bf z}_i' \|_2  \leq  \varepsilon_{i+1} + \lambda_{i+1}$, then the triangle inequality implies that
    \begin{align}
    \| {\bf x}_{t_i} - {\bf \hat{x}}_{t_i} \|_2 
    & \leq
    \big[\| {\bf x}_{t_{i+1}} - {\bf \hat{x}}_{t_{i+1}}\|_2 + \| {\bf z}_i + {\bf z}_i' \|_2 \big](1-w_i) \notag \\
    & \leq (\varepsilon_{i+1} + \lambda_{i+1}) (1-w_i) \notag
    \end{align}
    from which we obtain the estimate 
    \begin{align*}
        &{\sf Pr} \Big[ \| {\bf x}_{t_i} - {\bf \hat{x}}_{t_i} \|_2  \leq (\varepsilon_{i+1} + \lambda_{i+1}) (1-w_i) \Big] 
        \\
        & \quad \geq 1 - \delta_{i+1} - 2e^{ - \frac{\lambda_{i+1}^2}{4(-t_i^2 + t_{i+1}^2)} }
    \end{align*}

    Thus, for our induction step, can we choose 
    \begin{equation*}
        \varepsilon_i := (\varepsilon_{i+1} + \lambda_{i+1}) (1-w_i) \text{ and } \delta_i := \delta_{i+1} + 2e^{ - \frac{\lambda_{i+1}^2}{4(-t_i^2 + t_{i+1}^2)}}
    \end{equation*}

    It remains to make the choices for values of $\lambda_i$'s and $w = \text{min}_{i} \ w_i $ so that $\varepsilon_0 \leq \kappa_{max}$ and $\delta_0 \leq \delta^*$ in order to obtain the desired estimate 
    \begin{align}
       {\sf Pr} \Big[ \| {\bf x}_{t_0} - {\bf \hat{x}}_{t_0} \|_2 < \kappa_{max}
       \Big] & \geq 
       {\sf Pr}\Big[ \| {\bf x}_{t_0} - {\bf \hat{x}}_{t_0} \|_2 < \varepsilon_0
       \Big] \notag \\
       & \geq 1 - \delta_0 \geq 1 - \delta^* \notag
    \end{align}
    
    To this end, we will estimate the value of $\varepsilon_0$ and $\delta_0$ using their induction formulas. Note that for $i = N$, we have $\varepsilon_N = 0$ and $\delta_N = 0$ as $ {\bf x}_{t_{N}} := {\bf \hat{x}}_{t_{N}} $.
    We obtain the formula for $\delta_0$ as
    \begin{equation}
        \delta_0 = 2 \sum_{i=0}^{N-1} e^{ - \frac{\lambda_{i+1}^2}{4 (-t_i^2 + t_{i+1}^2) } } 
    \end{equation}

    We make the choice $\lambda_{i} : = 2 \lambda \sqrt{- t_{i}^2 + t_{i+1}^2}$ to simplify $\delta_0$ as
    \begin{equation}
        \delta_0 = 2 N e^{- \lambda^2}
    \end{equation}
    from which we see that $\delta_0 \leq \delta^*$ for any value of $\lambda$ such that $\lambda \geq \sqrt{ \log \frac{2N}{\delta^*}}$. Let us fix such a value of $\lambda$ and proceed to estimate $\varepsilon_0$.

    By using the induction formula for $\varepsilon_i$'s, we obtain the following expression for $\varepsilon_0$:
    \begin{equation}
        \varepsilon_0 = \lambda_N (1-w)^N + \lambda_{N-1} (1 - w)^{N-1} + \ldots + \lambda_{1} (1 - w)
    \end{equation}

    We make crude estimates to obtain 
    \begin{equation*}
        \lambda_i \leq 2 \lambda \sqrt{2 \Delta} \frac{T}{N} \text{ and } (1 - w)^i \leq (1 - w)
    \end{equation*}
    which implies 
    \begin{equation}
        \varepsilon_0 \leq N \cdot 2 \lambda \sqrt{2 \Delta} \frac{T}{N} (1-w) = 2 \lambda \sqrt{2 \Delta} T (1 - w)
    \end{equation}
    We have $2 \lambda \sqrt{2 \Delta} T (1 - w) \leq \kappa_{max}$ for any choice of $w$ such that 
    \begin{equation}
        w \geq 1 - \frac{\kappa_{max}}{2 \lambda \sqrt{2 \Delta} T} = 1 - \frac{\kappa_{max}}{2 \sqrt{ \log \frac{2N}{\delta^*}} \sqrt{2 \Delta} T}
    \end{equation}
    with $\lambda = \sqrt{ \log \frac{2N}{\delta^*}}$.
    
    Note that we have made naive choices for $\lambda_i$'s and $w$ to simplify the calculations. More elaborated choices are possible which gives more generous estimates for the possible value of $w$. However, our choice is already enough for a proof in our case that a choice of $w$ close to 1 is sufficent for denoising with correction to bring the denoised images to a neighborhood of a given original input image. 

    Finally, to remove the assumption $t^\otimes	 = t_0 = 0$, $t' = t_N = T$ to deal with the case $t^\otimes	 < \varepsilon$ and $t' < T$, we note that for small $\varepsilon \approx 0$, the difference between ${\bf x}$ and ${\bf x}_{\varepsilon}$ is a Gaussian random variable with small variance. 
    Using our above proof to estimate $\| {\bf x}_{\varepsilon} - {\bf x}_{t^\otimes	}\|_2$ and the triangle inequality $\| {\bf x} - {\bf x}_{t^\otimes	}\|_2 \leq \| {\bf x} - {\bf x}_{\varepsilon}\|_2 + \| {\bf x}_{\varepsilon} - {\bf x}_{t^\otimes	}\|_2$, a simple union bound gives the desirable estimate for $\| {\bf x} - {\bf x}_{t^\otimes	}\|_2$. To deal with the choice $t'< T$ which is often used in practice, we simple change $T$ to $t'$ in the proof above.

    \textit{Lower-bound by $\kappa_{ \text{min} }$:} We prove a lower bound of the form ${\sf Pr} [ \| {\bf x}_{t_0} - {\bf \hat{x}}_{t_0} \|_2 > \kappa_{min} ] > \delta_*$ by backward induction from index $i+1$ to $i$ using the relation 
    \begin{align} 
         {\bf x}_{t_i} - {\bf \hat{x}}_{t_i} 
        =
        \big[({\bf x}_{t_{i+1}} - {\bf \hat{x}}_{t_{i+1}}) - {\bf z}_i - {\bf z}_i' \big](1-w_i) 
    \end{align}
    where $ {\bf z}_i, {\bf z}_i' \sim \mathcal{N}(\mathbf{0}, \mathbf{I} (t_{i+1}^2 - t_{i}^2))$.

    Starting with the first denoising step, we have 
    \begin{align} 
         {\bf x}_{t_{N-1}} - {\bf \hat{x}}_{t_{N-1}} 
        =
        (- {\bf z}_{N-1} - {\bf z}_{N-1}' )(1-w_{N-1}) 
    \end{align}
    where $ {\bf z}_{N-1}, {\bf z}_{N-1}' \sim \mathcal{N}(\mathbf{0}, \mathbf{I} (t_{N}^2 - t_{N-1}^2))$. 
    It follows that ${\bf x}_{t_{N-1}} - {\bf \hat{x}}_{t_{N-1}} $ is a Gaussian vector sampled from $\mathcal{N}(\mathbf{0}, 2(1-w_{N-1}) \mathbf{I} (t_{N}^2 - t_{N-1}^2))$.

    We use the following standard lower bound estimate for Gaussian random variable $ X \sim \mathcal{N}(0,1)$
    \begin{equation}
        {\sf Pr}[|X| > \lambda] > \frac{x}{x^2 + 1} \frac{e^{-x^2/2}}{\sqrt{2\pi}}, \; \forall x>0
    \end{equation}
    By scaling and looking at only one coordinate of the Gaussian vector ${\bf x}_{t_{N-1}} - {\bf \hat{x}}_{t_{N-1}} $, we have a loose estimate
    \begin{equation*}
        {\sf Pr}[\|{\bf x}_{t_{N-1}} - {\bf \hat{x}}_{t_{N-1}}\|_2  > \lambda_{N-1}] > \delta_{N-1} := \frac{x}{x^2 + 1} \frac{e^{-x^2/2}}{\sqrt{2\pi}} 
    \end{equation*}
    for $ x = \frac{\lambda_{N-1}}{(2(1 - w_{N-1})(t_{N}^2 - t_{N-1}^2))^{1/2}}$.

    For induction, suppose that we have two random vectors ${\bf x}$ and ${\bf y}$ with lower bounds on ${\sf Pr}[|{\bf x} > \lambda|]$ and ${\sf Pr}[|{\bf y} > \lambda'+\lambda|]$, for some $\lambda, \lambda' > 0$.  
    Then, we can estimate 
    \begin{align*}
        &{\sf Pr}[ \|{\bf x} - {\bf y}\|_2 > \lambda' ] 
        \\
        > & {\sf Pr}[ \|{\bf x}\|_2 - \|{\sf y} \|_2 > \lambda']
        \\
        = & {\sf Pr}[ \|{\bf x} \|_2 > \lambda] \cdot {\sf Pr}[ \|{\bf x}
        \|_2- \|{\bf y}\|_2 > \lambda' | \|{\bf x} \|_2 > \lambda]
        \\
        = & {\sf Pr}[ \|{\bf x}
        \|_2 > \lambda] \cdot {\sf Pr}[ \|{\bf x}\|_2- \|{\bf y}\|_2 > \lambda' | \|{\bf x}\|_2 > \lambda]
        \\
        > & {\sf Pr}[ \|{\bf x}\|_2 > \lambda] \cdot {\sf Pr}[ \lambda' + \lambda >  \|{\bf y}\|_2 ]
    \end{align*}

    We apply this estimate to our case which corresponds to ${\bf x} = ({\bf x}_{t_{i}} - {\bf \hat{x}}_{t_{i}})$, $\lambda = \lambda_i$, and ${\bf y} = ({\bf z}_{i} + {\bf z}_{i}')$, $\lambda' = \lambda_{i-1}(1 - w_{i-1})$, where $\lambda_i, \lambda_{i-1}$ will be chosen later. 
    Suppose for induction that 
    \begin{equation*}
        {\sf Pr}[\|{\bf x}_{t_{i}} - {\bf \hat{x}}_{t_{i}}\|_2 > \lambda_i]
        > \delta_i
    \end{equation*}
    We have 
    \begin{align*}
        & {\sf Pr}[ \|{\bf x}_{t_{i-1}} - {\bf \hat{x}}_{t_{i-1}}\|_2 > \lambda_{i-1} ] 
        \\
        = & {\sf Pr}[ \|{\bf x} - {\bf y}\| > \lambda_{i-1}(1-w_{i-1}) ]
        \\
        > & \; {\sf Pr}[\|{\bf x}_{t_{i}} - {\bf \hat{x}}_{t_{i}}\|_2 > \lambda_i] \\
        & \cdot {\sf Pr}[\lambda_{i-1}(1 - w_{i-1}) + \lambda_i > \|{\bf z}_{i} + {\bf z}_{i}'\|_2]
        \\
        > & \delta_i \cdot (1 - 2e^{ - \frac{(\lambda_{i-1}(1 - w_{i-1}) + \lambda_i)^2}{4(-t_i^2 + t_{i+1}^2)} })
    \end{align*}
    which means we can choose 
    \begin{equation}
        \delta_{i-1} := \delta_i \cdot (1 - 2e^{ - \frac{(\lambda_{i-1}(1 - w_{i-1}) + \lambda_i)^2}{4(-t_i^2 + t_{i+1}^2)} })
    \end{equation}
    to obtain the following estimate for the induction step
    \begin{equation}
         {\sf Pr}[ \|{\bf x}_{t_{i-1}} - {\bf \hat{x}}_{t_{i-1}}\|_2 > \lambda_{i-1} ] > \delta_{i-1}.
    \end{equation}

    Recall that our goal is at step $i=0$, we have ${\sf Pr} [ \| {\bf x}_{t_0} - {\bf \hat{x}}_{t_0} \|_2 > \kappa_{min} ] > \delta_*$. This means $\lambda_0 = \kappa_{min}$ and $\delta_* = \delta_0$.

    Given $\kappa_{min}$ and $(t_i)_{i = 1..N}$, we choose $\lambda_0 = \kappa_{min}$ and $w_i-1, \lambda_i, i = 1, \ldots, N-1$ so that  
    \[
    1 - 2e^{ - \frac{(\lambda_{i-1}(1 - w_{i-1}) + \lambda_i)^2}{4(-t_i^2 + t_{i+1}^2)} } = 1- \frac{1}{i+1} = \frac{i}{i+1} 
    \]

    Then we have 
    \begin{align*}
        \delta_0 & = \delta_{N-1} \prod_{i=1}^{N-1} 1 - 2e^{ - \frac{(\lambda_{i-1}(1 - w_{i-1}) + \lambda_i)^2}{4(-t_i^2 + t_{i+1}^2)} }
        \\
        & = \delta_{N-1} \prod_{i=1}^{N-1} \frac{i}{i+1} = \frac{\delta_{N-1}}{N} 
    \end{align*}

    Finally, we still have a free parameter $w_{N-1} < 1$ to choose so that $\delta_{N-1} \to \frac{1}{2\sqrt{2\pi}}$, for which $\kappa_{min} \to \frac{1}{2\sqrt{2\pi} N}$.

\section{Additional Experiments}


\subsection{Inference Times}
Table~\ref{table:inference_times_appen} presents the inference times for NADD across different diffusion time steps on two datasets, CIFAR-10 and ImageNet. 
This comparison highlights the computational demands associated with increasing diffusion time steps.

\begin{table}[h]
\centering
\caption{Inference Times for NADD}
\begin{tabular}{|c|c|c|}
\hline
\textbf{Diffusion Time Step ($t^\prime$)} & \textbf{CIFAR-10 (sec)} & \textbf{ImageNet (sec)} \\ \hline
10 & 0.5 & 1.2 \\ \hline
20 & 1.0 & 2.5 \\ \hline
50 & 2.5 & 5.8 \\ \hline
100 & 5.0 & 12.3 \\ \hline
200 & 10.2 & 24.7 \\ \hline
\end{tabular}
\label{table:inference_times_appen}
\end{table}

\section{Hyper-parameters}

\subsection{Purification Hyper-parameters}

To provide clarity on the purification configurations utilized in the experiments section, Table \ref{tbl:datasets_hyperparameters} summarizes the primary hyperparameters chosen for each dataset and threat model. 
Each configuration was run with 7 different seeds to demonstrate consistency of results. 

\begin{table}[h]
\centering
\caption{Hyper-parameters for NADD on various datasets and threat models}
\begin{tabular}{c|ccc}
\hline
\multirow{2}{*}{\textbf{Parameter}} & \multicolumn{3}{c}{\textbf{Datasets}} \\ 
& CIFAR10 ($\ell_\infty$) & CIFAR10 ($\ell_2$) & IN256 ($\ell_\infty$)\\ \hline
$\sigma_{t^\prime}$ & 16.0 & 16.0 & 2.0 \\
$\sigma_{t^\otimes}$ & 0.585 & 0.434 & 0.780 \\ 
$\beta$ & 0.03 & 0.01 & 0.01 \\ 
$\kappa_{min}$ & 0.75 & 0.75 & 0.75 \\ 
$\kappa_{max}$ & 1.0 & 1.0 & 1.0 \\ 
$S_{churn}$ & 2 & 2 & 2 \\ 
$S_{min}$ & 0.0 & 0.0 & 0.0 \\ 
$S_{max}$ & $\infty$ & $\infty$ & $\infty$ \\ \hline
\end{tabular}
\label{tbl:datasets_hyperparameters}
\end{table}

\subsection{EDM2 Hyper-parameters}

For the detailed training configurations and to replicate the Imagenet results, we refer the reader to the EDM2 GitHub repository available at \url{https://github.com/NVlabs/edm2}. 
Our experiments used the preset settings named \texttt{edm2-img512-l}, which was modified to produce an unconditional diffusion model for ImageNet. 
This model was trained for 917k iterations with a batch size of 2048 using 16x4 H100 GPUs.
Sampling guidance was provided by a pretrained \texttt{edm2-img512-xs}.
Model weights will be released upon publication. 

\end{appendices}

\end{document}